\documentstyle[12pt]{article}

\begin{document}


\makeatletter
\renewcommand{\@biblabel}[1]{\quad#1.}
\makeatother

\hsize=6.15in
\vsize=8.2in
\hoffset=-0.42in
\voffset=-0.3435in

\normalbaselineskip=24pt\normalbaselines

\begin{center}
Published in:  {\it PLoS Comput Biol} {\bf 11(10)}: e1004532
\end{center}

\vspace{0.45cm}

\begin{center}
{\large \bf Cortical composition hierarchy driven by spine proportion economical
maximization or wire volume minimization }
\end{center}

\vspace{0.15cm}

\begin{center}
{Jan Karbowski}
\end{center}

\vspace{0.05cm}

\begin{center}
{\it   Institute of Applied Mathematics and Mechanics,  \\
University of Warsaw, ul. Banacha 2, 02-097 Warsaw, Poland;  } 
\end{center}


\vspace{0.1cm}

\begin{abstract}
The structure and quantitative composition of the cerebral cortex are 
interrelated with its computational capacity. Empirical data analyzed here 
indicate a certain hierarchy in local cortical composition. Specifically,
neural wire, i.e., axons and dendrites take each about 1/3 of cortical
space, spines and glia/astrocytes occupy each about $(1/3)^{2}$, and
capillaries around $(1/3)^{4}$. Moreover, data analysis across species 
reveals that these fractions are roughly brain size independent, which 
suggests that they could be in some sense optimal and thus important for 
brain function. Is there any principle that sets them in this invariant way? 
This study first builds a model of local circuit in which neural wire, spines,
astrocytes, and capillaries are mutually coupled elements and are treated
within a single mathematical framework. Next, various forms of wire minimization 
rule (wire length, surface area, volume, or conduction delays) are analyzed, of
which, only minimization of wire volume provides realistic results that are 
very close to the empirical cortical fractions. As an alternative, a new principle 
called ``spine economy maximization'' is proposed and investigated, which is 
associated with maximization of spine proportion in the cortex per spine size 
that yields equally good but more robust results. Additionally, a combination of 
wire cost and spine economy notions is considered as a meta-principle, and it 
is found that this proposition gives only marginally better results than either 
pure wire volume minimization or pure spine economy maximization, but only if
spine economy component dominates. However, such a combined meta-principle 
yields much better results than the constraints related solely 
to minimization of wire length, wire surface area, and conduction delays. 
Interestingly, the type of spine size distribution also plays a role, and 
better agreement with the data is achieved for distributions with long tails. 
In sum, these results suggest that for the efficiency of local circuits wire 
volume may be more primary variable than wire length or temporal delays, and 
moreover, the new spine economy principle may be important for brain evolutionary 
design in a broader context.

\end{abstract}





\vspace{0.1cm}

\noindent Email:  jkarbowski@mimuw.edu.pl


\vspace{2.3cm}


\noindent {\bf Author Summary:}

\vspace{0.4cm}

Cerebral cortex is an outer layer of the brain in mammals, and
it plays a critical part in various cognitive processes such as  
learning, memory, attention, language, and consciousness.
The cerebral cortex contains a number of neuroanatomical parameters 
whose values are essentially conserved across species and brain sizes,
which suggests that these particular parameters are somehow important 
for brain efficient functioning. This study shows that the fractional 
volumes of five major cortical components both neuronal and non-neuronal
(axons, dendrites, spines, glia/astrocytes, capillaries) are also 
approximately conserved across mammals, and neural wire (axons and 
dendrites) occupies the most of cortical space. Moreover, the fractional
volumes form a special hierarchy of dependencies, being approximately equal 
to integer powers of 1/3. Is there any evolutionary principle of cortical 
organization that would explain these properties? This study finds that 
there are two different theoretical principles that can provide answers: 
one standard related to minimization of neural wire fractional volume, and 
a new proposition associated with economical maximization of spine content. 
However, the latter principle produces more robust results, which suggests
that spine economical maximization is potentially an alternative to the 
more common ``wire minimization'' in explaining the cortical layout. 
Therefore, the current study can become an important contribution to our 
understanding (or debating) of the main factors influencing 
the evolution of local cortical circuits in the brain.

\newpage

{\large \bf Introduction}

The gray matter of cerebral cortex is composed primarily of neurons with
their extended axonal and dendritic processes, synapses (mostly dendritic spines) 
connecting different neurons, non-neuronal cells called glia (of which astrocytes 
are likely the most important), and microvasculature (capillaries). It is thought 
that neurons and synapses are the main computational and functional elements, 
whereas glia and capillaries serve supporting or modulatory roles associated with 
supplying metabolic substrates (oxygen and glucose) to neurons and synapses
based on their demands \cite{iadecola,attwell2010}. The experimental data show 
that there are certain scaling regularities in the arrangement of neuronal,
synaptic \cite{wen2009,snider,cuntz}, and capillary \cite{karbowski2011} 
processes. It is commonly hypothesized that these scaling rules may be the 
consequence of the design principle called neural ``wiring minimization'' or 
``wiring economy'' \cite{wen2009,cherniak,chklovskii,wen2005,klyachko}, or efficiency 
in functionality-cost trade-offs \cite{budd2010,budd2012,bullmore,karbowski2014}. 
However, there are some indications that wiring length minimization is not
enough to explain the pattern of global connections in the macaque cortex and 
in the nematode {\it C. elegans} nervous system \cite{kaiser,chen}, and even 
small-scale networks of macaque and cat visual cortices perform sub-optimally
in terms of wire length reduction \cite{cherniak2004}. This opens 
a possibility for some other optimization principles governing brain architecture.

Here, it is proposed a new, alternative, design principle that can be called 
``spine economical maximization'' or spine economy. It is based on maximization
of spine proportion in the cortex with simultaneous penalization of spine size. 
This scenario is equivalent to requiring that in a given cortical volume there are 
as many information storing connections between neurons as possible (maximal
functionality) that cost as little energy as possible (economical approach, for 
bigger spines use more energy). 
Moreover, we require that these connections are efficiently coupled with astrocytes 
and capillaries, and the whole system of neuronal and non-neuronal elements is 
treated within a single theoretical framework. The choice of the guiding principle 
associated with spine economy is motivated by the following empirical results:
(i) Axo-spinal synaptic connections are the most numerous in the cortical gray
matter \cite{zecevic} and are very important not only for short-term neuronal 
communication \cite{budd2012}, but also for long-term information storage 
\cite{kasai}; (ii) Mammalian brain is metabolically expensive 
\cite{aiello,attwell2001,karbowski2007}, 
which means that energy is an important constraint on brain function 
\cite{karbowski2014,levy,laughlin}; 
(iii) Spines are likely the major users of metabolic energy in the cortical gray 
matter \cite{alle,harris,karbowski2012}, as is reflected in a strong correlation
between cortical synaptogenesis and its energetics \cite{karbowski2012,chugani}, 
which suggests that spine number or size can be limited by available energy.

The new hypothetical principle associated with spine economy is tested here for 
local organization of cortical circuits. Specifically, it is tested if fractional 
occupancy of space by the main cortical components predicted on a basis of the 
spine economical maximization principle agrees with empirical data. Fractional 
distribution of volume taken by neuronal and non-neuronal elements should be an 
important aspect of local cortical organization, because densities of neurons, glia, 
and vasculature are mutually correlated across cortical regions and layers 
\cite{tsai,weber,mccaslin}. Thus, too much space taken by one component
can lead, due to competition, to underperformance of other components \cite{montague} 
or to an excessive cortical size \cite{kaas}, both of which can be undesirable 
for brain efficient design and functionality. Thus, some neuroanatomical balance 
between fractional volumes of cortical elements seems necessary. Unfortunately, 
it is virtually not known whether these fractions are variable or preserved across 
species. This interesting topic was only briefly addressed before, with the 
suggestion that the combined fraction of neuronal wire (axons and dendrites) can 
result from minimization of temporal delays in inter-neuronal signaling 
\cite{chklovskii}. However, from an evolutionary perspective, the knowledge of 
fractional distribution of all major cortical components, also those supplying 
metabolic energy, should add an important information to our understanding of 
the geometric layout of the cortex and for testing various hypotheses concerning 
its design principles \cite{bassett}.

The paper is organized as follows. First, empirical data on fractional volumes
of cortical components are analyzed. In particular, we look for regularities 
in the data within and across species. We build and study a theoretical model of 
cortical composition with coupled neuronal and non-neuronal elements. Next, we
investigate which optimization principle can best explain the empirical facts. Three 
classes of optimization models are considered. One is based on a standard principle of 
neural wire minimization and includes minimization of wire length, wire surface
area, wire volume, and local temporal delays. Second class is based on a new
proposition of spine economical maximization. Third class is a linear combination 
of the first two types of models, i.e., it mixes wire cost with spine economy. 
All kinds of models are based on an implicit assumption that evolution had optimized 
the nervous system according to some rules 
\cite{bullmore,karbowski2014,parker,alexander,striedter}.

\vspace{1cm}


{\large \bf Results}

\vspace{0.3cm}

\noindent
{\bf Empirical composition of the cerebral cortex across mammals
reveals hierarchical organization.} \\
Existing experimental data on fractional volumes of cortical gray matter components
were analyzed (see the Methods), and it is found that these fractions exhibit 
a certain hierarchy, since they can be approximated by integer powers of 1/3
(Table 1; Fig. 1). Specifically, axons and dendrites occupy each about 1/3 
of cortical space, dendritic spines and glia/astrocytes constitute each roughly
$(1/3)^{2}$ of the cortex, and capillaries take an extremely small volume fraction
around $(1/3)^{4}$ (Table 1; Fig. 1). This regularity is called here the rule
of ``powers of 1/3''. Moreover, an allometric analysis reveals that the fractions 
of all examined cortical components are species- and brain size independent, i.e. they
do not correlate significantly with cortical size and scale with exponents close
to zero (Table 1; Fig. 2). Typical values for axons: exponent=$-0.036$,
$R^{2}=0.083$, $p=0.713$; for dendrites: exponent=$-0.002$, $R^{2}=0.037$, $p=0.717$;
for spines: exponent=$-0.013$, $R^{2}=0.008$, $p=0.885$; for glia/astrocytes:
exponent=0.031, $R^{2}=0.180$, $p=0.477$; and for capillaries: exponent=0.064,
$R^{2}=0.243$, $p=0.399$ (Fig. 2).

\vspace{0.35cm}

\noindent
{\bf Optimality principles for local cortical circuits.} \\
It is possible that the relative constancy of the fractional occupancy of cortical 
space and its simple hierarchy reflect some kind of evolutionary optimized principle. 
Three such principles are considered: one associated with neural wire minimization,
second with spine proportion economical maximization, and the third proposition is
the mixture of the first two. The last choice means that we consider the possibility
of some ``meta-principle'', which includes the contributions of both primary principles 
(wire minimization and spine economy maximization) with some weights related to their 
importance. The most general fitness function $F$, or benefit-cost function, associated 
with such a meta-principle, which we want to minimize, is given by (see the Methods, in
particular the section ``The fitness functions'' for the derivation)

\begin{eqnarray}
F= f\frac{(rx + y)}{\overline{u}^{\gamma_{1}}} - (1-f)\frac{s}{\overline{u}^{\gamma_{2}}} 
+ \lambda\left(x + y + s + g + c - 1 \right),
\end{eqnarray}\\
where $x, y, s, g, c$ are respectively volume fractions of axons, dendrites, spines, 
glia, and capillaries in the cortical gray matter, and $\lambda$ is the Lagrange
multiplier associated with a mathematical constraint of the fractions normalization. The symbol 
$f$ is the control parameter $(0 \le f \le 1)$, or mixing ratio, measuring a relative 
contribution (importance) to the fitness function of the two contrasting notions: wire 
minimization and spine economical maximization. If $f=1$ then the function $F$ corresponds 
to wire cost only, whereas if $f=0$ then the function $F$ describes spine economy only 
(the negative sign in front of $(1-f)$ is necessary to obtain maximum for spine content). 
The symbol $\overline{u}$ is the average spine volume, $r$ is some 
positive measure of asymmetry between axons and dendrites, $\gamma_{1}$ and $\gamma_{2}$ 
are positive exponents. Different values of $\gamma_{1}$ correspond to different kinds 
of wire minimization. In particular, $\gamma_{1}= 0$ relates to wire volume minimization,
$\gamma_{1}= 1/3$ corresponds to wire surface area minimization, $\gamma_{1}= 2/3$ is
associated with wire length minimization, and $\gamma_{1}= 5/6$ relates to local temporal 
delays minimization (see the Methods). The parameters $r$, $\gamma_{2}$, and $f$ constitute 
the free parameters. 

In the next sections we study theoretical consequences of minimization of the fitness
function represented by Eq. (1). Specifically, we find optimal values of fractional
volumes of the cortical components and compare them with the empirical data.

\vspace{0.35cm}

\noindent
{\bf Optimal fractional volumes of cortical components: wire minimization
vs. spine economical maximization.} \\
Wire minimization principle corresponds to $f=1$ in Eq. (1). For this scenario the optimal 
fractional volumes of cortical components depend on two parameters: $\gamma_{1}$ and $r$.
The dependence of optimal fractions on $\gamma_{1}$ is critical. If $\gamma_{1}= 0$ 
(the case of wire fractional volume minimization) then the theoretical fractions can be 
similar to the empirical values in Table 1 (the next-to-last line) only when $r\sim 1$ 
(Fig. 3). If  $\gamma_{1} > 0$ (the other types of wire minimization), regardless of 
the value of $r$, the optimal fractions of glia/astrocytes and capillaries vanish, 
which is unrealistic (Fig. 3A, C). Moreover, only the case $\gamma_{1}=0$ produces finite 
reliable values of the average spine volume $\overline{u}$ (Table 2); for $\gamma_{1} > 0$ 
we obtain $\overline{u}\mapsto \infty$. The abrupt transition in solutions from $\gamma_{1}= 0$ 
to $\gamma_{1} > 0$ is reminiscent of discontinuous phase transitions. The dependence 
of the optimal fractional volumes on $r$ is more smooth (if $\gamma_{1}= 0$; Fig. 3B,D). 
Realistic fractions are obtained for almost symmetrical situation, i.e. when $r$ is close 
to 1; too small or too large $r$ produces vanishing fractions of either dendrites or axons 
(Fig. 3B, D). Moreover, all the described results look qualitatively similar regardless of 
the asymptotic nature of the distributions of spine sizes (compare Figs. 3A,B vs. Figs. 3C,D).

Alternatively, for spine economical maximization principle, i.e. for $f=0$ in Eq. (1), 
the optimal fractions of cortical components depend moderately on the exponent $\gamma_{2}$. 
This dependence is displayed in Fig. 4 for four distributions of spine sizes: 
two with short tails (Gamma and Exponential distributions) and two with long tails 
(Log-logistic and Log-normal). As can be seen the optimal fractions of axons and dendrites 
are equal and about 0.4, and they are qualitatively almost independent of $\gamma_{2}$ 
and spine size distribution. Similarly, the capillary content is relatively stable at about 
0.01 or a little less (Fig. 4). In contrast, the fractions of spines and astrocytes/glia 
can change significantly as a function of $\gamma_{2}$. Typically, the spine proportion
decreases with increasing $\gamma_{2}$, whereas glia/astrocytes content depends 
non-monotonically on $\gamma_{2}$, but both of them are restricted from above
(spines by $\sim 0.16$ and astrocytes by $\sim 0.14$). These trends are preserved 
as we change the distributions of spine volumes from short-range to long-range 
(Fig 4A,B vs. Fig 4C,D). The pattern for the remaining two distributions (Gamma n=1 
and Rayleigh; not shown) is very similar. Overall, the optimal fractions of cortical 
components can either vary with $\gamma_{2}$ or not, but these dependencies are
essentially invariant with respect to the distribution type of spine sizes. More 
importantly, the optimal fractional volumes are quantitatively similar to the empirical 
values given in Table 1 (the next-to-last line) for a broad range of $\gamma_{2}$.

The degree of similarity between the theory and the data is quantified by two
different measures, their Euclidean distance ED, and their Mahalanobis distance MD 
(see Eqs. 34 and 35). The latter is more general, as it takes variability in the data
into account.  Both measures yield qualitatively very similar results (compare
Figs. 5 and 6). For wire minimization principle, ED and MD depend biphasically 
(non-monotonically) on $r$ if $\gamma_{1}= 0$ in a similar fashion for 
all distributions of spine volumes, with characteristic sharp minima for $r\sim 0.95$, 
for which there are the best matches to the data (Figs. 5A, 6A). The overall best
results are achieved for Log-normal distribution (ED=0.010, MD=1.405), and other
distributions especially those with short-tails produce higher ED and MD (Table 2). 
Outside the optimal value of $r$ the values of ED and MD grow rather fast, and the 
agreement between theory and the experiment becomes weak (Figs. 5A, 6A). If however,
$\gamma_{1} > 0$, then ED and MD are both constant and relatively large 
(ED=0.15, MD=18.5), regardless of other parameters and distribution types 
(Figs. 5B, 6B), which implies that similarity with the data is always very 
weak in this case.

On the other hand, for spine economical maximization ED and MD exhibit minima for 
$\gamma_{2} < 1$ (Figs. 5C, 6C). Interestingly, neither the minimal values of ED and 
MD, nor associated with them the optimal values of $\gamma_{2}$ vary much as we change 
the spine distribution type for a given threshold $\theta$ (Figs. 5C, 6C). Specifically, 
the optimal $\gamma_{2}$ is in the range $0.15-0.40$ for $\theta=0.100$ $\mu$m$^{3}$, 
and  $0.35-0.75$ for $\theta=0.321$ $\mu$m$^{3}$ (Table 3). The minima of ED and MD 
for heavy-tailed distributions are much broader than for short-tailed, which suggests 
that heavy-tailed distributions can be more flexible in comparison to the data (Figs. 5C,
6C). This follows from the fact that heavy-tailed distributions depend on an additional 
parameter that can be adjusted (Fig. A in Supporting S1 Text). Moreover, the best optimal 
cortical fractional volumes in Table 3 look very similar across different distributions of
spine sizes. ED values for these best solutions are also very similar (Table 3). 
The smallest possible minimal value of ED is 0.038, and it is reached by four different 
distributions, independent of the asymptotic tail (Gamma n=2, Rayleigh, Log-logistic, 
and Log-normal). (As a comparison, for a hypothetical situation when all theoretical 
optimal cortical components were uniformly distributed, i.e. each of them were 0.2, 
we would obtain ED=0.342, which is about 10 times larger than the actual minimal ED 
in Table 3). The situation is different for the MD distance. Despite a high degree of
similarity among the optimal fractional volumes across different distributions, MD 
discriminates these cases much better than ED (Table 3). Distributions having identical 
ED have in general different MD values (Table 3). The overall trend is such that
the best fits to the empirical fractional volumes are given by the distributions
of spine sizes with long-tails. The smallest possible minimal MD is achieved for
Log-normal distribution (MD=1.788; Table 3). Because of its higher generality, only
the MD measure is used subsequently for comparing theoretical predictions with
the data.

The degree of the sensitivity of the optimal cortical fractions on the threshold 
$\theta$ for spine formation is qualitatively mostly similar for both principles of 
cortical organization considered here (Figs. 5B,D and Figs. 6B,D), except for the 
case of wire minimization with $\gamma_{1} > 0$, for which ED, MD, and cortical 
fractions are independent of $\theta$ but are not realistic (Figs. 5B and 6B). 
Indeed, for wire volume minimization (with $\gamma_{1}= 0$) and spine economical 
maximization, ED and MD depend very strongly on $\theta$ if $\theta  < 0.2$ 
$\mu$m$^{3}$ (Figs. 5B,D and 6B,D). However, if $\theta > 0.2$ $\mu$m$^{3}$ 
this dependence is milder in both cases (Figs. 5B,D and 6B,D).

\newpage

\noindent
{\bf Optimal spine sizes and probability of spine formation: wire
minimization vs. spine economical maximization.}  \\
Optimization of the fitness function $F$ also yields optimal average spine volume 
$\overline{u}$ and indirectly the conditional probability of spine formation $P$ 
(see the Methods). We consider the results for pure wire minimization and pure spine
economy maximization.

As was previously noted, for wire minimization principle ($f=1$ in Eq. 1) with 
$\gamma_{1} > 0$, we obtain $\overline{u}\mapsto \infty$, and consequently $P=1$, both 
of which are unrealistic. For wire minimization with $\gamma_{1}=0$ (wire volume
minimization), both $\overline{u}$ and $P$ depend non-monotonically on the parameter 
$r$ (Fig. 7A, B). For short-range distributions of spine size, $\overline{u}$ and $P$ 
are positively correlated, whereas for the distributions with heavy-tails these two 
quantities are anti-correlated. Thus for wire minimization there is no clear 
one-to-one correspondence between average spine size and conditional probability of 
spine formation. Among all the distributions, for the optimal value of $r$ ($r\approx 0.95$), 
the heavy-tailed Log-normal produces the most realistic spine volume 
$\overline{u}= 0.24-0.56$ $\mu$m$^{3}$, i.e. the closest to the empirical values 
($0.2-0.4$ $\mu$m$^{3}$ for human and macaque monkey \cite{benavides,villalba}), 
regardless of the value of threshold $\theta$ (Table 2; Fig. 7A). For short-tail 
distributions the values of $\overline{u}$ are strongly threshold $\theta$-dependent 
(Table 2).

In contrast, for spine economical maximization ($f=0$ in Eq. 1) the quantities $\overline{u}$ 
and $P$ depend monotonically on the exponent $\gamma_{2}$, and thus are positively correlated 
for all distributions, i.e. small spine volumes are generally associated with low 
probabilities $P$ and vice versa (Fig. 7C,D). For optimal values of $\gamma_{2}$ ($ < 1$) 
the average spine volumes have very similar values regardless of the spine size distribution 
and the threshold $\theta$ (Gamma n=2 and Rayleigh distributions are exceptions), 
typically: $\overline{u}= 0.44-0.67$ $\mu$m$^{3}$ (Table 3), which is also close to
the experimental data \cite{benavides,villalba}.

\vspace{0.35cm}

\noindent
{\bf Optimal fractional volumes of cortical components: combined
``wire min and spine max'' meta-principle.}  \\
Now we consider the scenario when wire minimization and spine economical maximization 
notions are mixed together simultaneously, and optimal fractional volumes are obtained 
by minimization of the meta fitness function in Eq. (1) with the control parameter 
$0 < f < 1$. In general, the best results in terms of agreement with the empirical data 
are reached for the cases when spine economy rule dominates in the meta fitness function,
i.e. $f \ll 1$, (Fig. 8; Tables 4-6). More precisely, MD distance has very shallow minima
in the range $f \sim 0.1-0.3$ (Fig. 8). The worst results are found for the cases when wire 
minimization rule prevails ($f\mapsto 1$), and intermediate results are obtained for the 
balanced case when spine economy max and wire min are treated with equal weights, i.e. 
$f\sim 0.5$ (Fig. 8 and Tables 4-6). However, there is an exception to this tendency, 
namely, the mixture of wire volume minimization and spine economy maximization, which 
yields invariant results and good similarity with the data regardless of the value of 
mixing ratio $f$ (Fig. 8C and Tables 4-6).

For the case when spine economy predominates ($f=0.1$) in $F$, we obtain that mixing 
spine max with a small fraction of any type of wire min rule (arbitrary $\gamma_{1}$) 
gives essentially the same results for MD (Table 4), and they are very similar to the results 
for pure wire volume minimization (Table 2) and pure spine economy maximization (Table 3). 
This means that for this particular scenario all kinds of wire minimization rule, 
such as length min, surface area min, volume min, or delays min, are equally reasonable,
but they are less important than spine economy.
This, however, is not the case when wire minimization prevails in the mixing of notions, 
or when there is a balance between them (cases $f=0.9$ and $f=0.5$, respectively). 
In these scenarios, wire volume min (and occasionally wire surface min) gives noticeably 
better results than either wire length min or delays min (Tables 5 and 6).
However, it has to be stressed that mixing together spine economy rule with any of the 
wire cost rules with $\gamma_{1} > 0$ (wire length, wire surface, delays), at any ratio,
always gives better results than for these wire cost rules alone, i.e., for $f=1$
(Fig. 8A,B,D).

Another interesting feature of the results in Tables 4-6 is that the best agreements 
with the data are almost always recorded for the distributions of spine sizes with long tails. 
This trend is conserved for a large spectrum of the values of $f$, however, as $f\mapsto 1$ 
(the notion of wire min dominates in $F$) the similarity to the data becomes weaker or 
even non-existent (see Fig. 8 for Log-normal distribution).

To summarize, mixing the rules of spine economy max with wire min does not give significantly 
better results than for pure spine economy max or pure wire volume min. At best, such 
a mixing yields marginally better results, but only if the spine component prevails 
($f \ll 1$) in the meta fitness function $F$.

\newpage

{\large \bf Discussion}

\vspace{0.3cm}

\noindent
{\bf Hierarchy in cortical composition and the rule ``powers of 1/3''.} \\
Experimental data analyzed in this paper indicate that the distribution of the 
basic components in the gray matter of cerebral cortex is relatively stable across 
mammalian species (Table 1; Fig. 2). Axons and dendrites occupy on average similar 
fractions $\sim 1/3$ of cortex volume, spines and glia/astrocytes take each about 
1/10 of the cortex, and capillaries constitute roughly $\sim 1/100$ of the cortical 
space. These numbers in themselves are interesting, because they form a special hierarchy 
of dependencies, being approximately integer powers of 1/3 (Fig. 1). Specifically, 
the content of spines or glia/astrocytes is roughly equal to the square of axon 
(or dendrite) content, and capillary content in turn is approximately equal to the 
square of spine content (Table 1). This component hierarchy may be somehow important 
for efficient cortical computation, and thus worth investigating.

\vspace{0.35cm}

\noindent
{\bf Summary of the main theoretical results: the importance of spine economy principle.} \\
Hierarchy in cortical composition was a motivation for theoretical considerations
about principles governing organization of the cerebral cortex. The goal was to
provide a theoretical explanation of this hierarchy starting from a neurobiologically 
plausible yet simple principle. Two basic principles are considered: different forms of 
a standard neural wire minimization \cite{wen2009,cherniak,chklovskii,wen2005,stevens},
and the new one proposed here called spine economical maximization. The latter rule
is related to maximization of spine content in the cortex with simultaneous minimization
of average spine size (which is supposed to reduce the metabolic cost). We also study
a mixture of the two principles as a meta principle. The optimal outcomes of these
models are compared to the experimental data, using two similarity measures (Euclidean
and Mahalanobis distances).

This study shows that from many implementations of wire minimization concept, only the 
minimization of wire fractional volume ($\gamma_{1}= 0$) can give reasonable results 
that are close to the experimental data, if a free parameter ($r$ in Eq. 21) is 
chosen appropriately (compare Tables 1 and 2; Figs. 5A and 6A). The other possibilities 
related to wire minimization (with $\gamma_{1} > 0$), such as minimization of wire length,
its area, or conduction delays, yield unrealistic results: zeros for volume fractions 
of glia/astrocytes and capillaries (Fig. 3A,C), resulting in relatively high values
of ED and MD (Figs. 5B and 6B), and infinite values of the average spine volume, which is 
clearly wrong. The last result follows from the fact that the minimal value of $F_{w}$ 
in Eq. (21) is precisely zero, which takes place only for $\overline{u}= \infty$ 
(other possibility with vanishing of axonal $x$ and dendritic $y$ fractions is 
forbidden, since that would imply that all fractions are zero, which would violate 
the fractions normalization constraint represented by Eq. (31)).

On the other hand, the principle of spine economic maximization produces the fractional 
cortical volumes that are also close to the data (compare Tables 1 and 3), but they 
do not require such a careful tuning of a free parameter ($\gamma_{2}$ in Eq. 24), 
especially for the distributions of spine volume with heavy-tails (compare the scales in 
Figs. 5C vs. 5A and in Figs. 6C vs. 6A). Thus, both principles provide quantitatively 
similar results, however, maximization of the simple benefit-cost function $F_{s}$ with 
spine content in the centerpiece (see Eq. 24) produces a more robust outcome. This conclusion 
is consistent with suggestions that neural systems are not exclusively optimized for 
minimal wiring length or component placement \cite{kaiser,chen}, and other factors or their 
combinations can be also involved \cite{budd2010,budd2012,bullmore,karbowski2014,kaiser,chen}. 
This suggests that economic spine content maximization can possibly provide an additional 
and/or alternative mechanism that is used by evolution to regulate the efficiency of cortical 
circuits.

As a third possibility we consider the meta principle that combines spine economy with 
wiring cost, with some mixing ratio $f$. This mixing scenario improves greatly the results
associated with minimization of wire length, wire surface area, and temporal delays
($\gamma_{1} > 0$), by making MD distances much smaller, but only when the spine economy 
dominates in the mixing of notions in the fitness function (Eq. 1). Thus, clearly the 
presence of the principle of spine economical maximization is necessary to make the concepts 
of wire length and temporal delays minimizations to be relevant candidates for the 
explanation of cortical composition hierarchy (Table 4). However, it must be emphasized
that mixing of wire volume min ($\gamma_{1}= 0$) with spine economy max does not
produce noticeably better results (only a tiny improvement) in comparison to the cases
when these two principles act in isolation (Fig. 8C).

Finally, for all three scenarios (either $f=0$, $f=1$, or $f$ between 0 and 1 in Eq. 1),
the best theoretical fractional volumes are generally obtained for distributions of spine 
volume with heavy-tails (Tables 2-6). In addition, many of these best solutions give
optimal average spine volumes $\overline{u}$, which despite some variability, are in good 
agreement with experimental values, which are 0.35 $\mu$m$^{3}$ for human \cite{benavides}, 
and 0.3-0.4 $\mu$m$^{3}$ for macaque monkey \cite{villalba}. Spine size correlates
with synaptic weight \cite{kasai}, and hence variability in spine size that is observed
among the best solutions is a positive feature, because it implies variability in
synaptic weights, which in turn is necessary for brain function.

The models for pure wire volume minimization and for pure spine economy differ
strongly in their predictions regarding the relationship between average spine
size and conditional probability of spine formation $P$ (Fig. 7A,B vs. Fig. 7C,D).
While for wire volume minimization there is no clear one-to-one correspondence,
for spine economy we obtain that small spines are associated with small $P$ (Fig. 7C,D). 
This means that for the latter principle very small spines are unlikely to form 
and thus are highly stochastic, even when an axon and dendrite are very close to 
each other. In contrast, large spines with sufficiently large energy capacity have 
high probability of forming stable synapses. Mathematically, this effect is achieved 
in the model by introducing the threshold $\theta$ for spine volume (see Eq. 3 in 
the Methods). This theoretical result for spine economical maximization is in line 
with experimental observations, indicating high stochastic motility of small spines 
and structural stability of larger ones \cite{kasai,bonhoeffer,holtmaat,statman,meyer}. 

\vspace{0.35cm}

\noindent
{\bf Previous directly related work.} \\
In the past there was only one directly related work associated with cortical 
composition and its theoretical basis \cite{chklovskii}. In that study, the 
authors analyze only the combined optimal fractional volume of axons and dendrites 
(called wire), which turns out to be close to the empirical value. Chklovskii 
et al \cite{chklovskii} used specific ``thought experiments'' to demonstrate
that the optimal wire fraction can be derived from several equivalent principles,
such as minimization of conduction delays, minimization of wiring length, 
and maximization of synaptic density. The present study is similar in
spirit, i.e. in the expectation that fractional volumes are the result of
some evolutionary optimization, but differs in the scope and details of what
is actually optimized. Apart from considering different forms of wire minimization,
we also investigate a new principle of spine economy, and its combination with
wire cost (meta principle represented by Eq. 1), and all three are analyzed in much 
more detail than in \cite{chklovskii}. In particular, in this work we study five 
cortical components both neuronal and non-neuronal, in contrast to \cite{chklovskii}, 
who considered only a simple division wire vs. non-wire. Moreover, we provide explicit 
formulae for fractional volumes of different components and their mutual couplings, 
based on a concept of geometric probability, efficient transport for astrocytes, 
and some neuroanatomical observations.

\vspace{0.35cm}

\noindent
{\bf Other formulations of the problem related to spine economy.} \\
One may wonder why in the fitness function related to spine economy (Eq. 24), the spine 
proportion ($s$) is maximized instead of (maybe more natural) numerical spine density 
($\rho_{s}$)? In fact, both possibilities are included in the fitness function, 
since these two quantities are linearly related ($s=\rho_{s}\overline{u}$). Thus the 
fitness function (Eq. 24) can be equivalently written as 
$F_{s} \sim \rho_{s}/\overline{u}^{\gamma_{2}-1}$, with the renormalized power of 
$\overline{u}$. However, since the best results are obtained for $\gamma_{2} < 1$, this 
new formulation implies that spine density and average spine volume both would have to 
be maximized, i.e., there would be no penalty on spine size. Thus, the original fitness 
function with spine proportion $s$ seems to better capture the energy constraint, which 
suggests that $s$ is a more primary variable than $\rho_{s}$.

\vspace{0.35cm}

\noindent
{\bf Organization of local vs global circuits.} \\
This study considers local cortical circuits, presumably corresponding to a cortical
column in size, and does not address the large-scale organization of the mammalian 
brain. In a present mathematical formulation of local circuits the notion of spatial
scale is not taken explicitly into account. This is justified by the fact that locality 
means that all elements are sufficiently close to each other so that there should not be
large detrimental delays in their communication. That temporal delays are not critical 
in local networks follows from a fact that diameters of intracortical axons and 
dendrites, directly related to velocities of propagated signals along them \cite{hursh}, 
are invariant with respect to brain size \cite{karbowski2014}. The situation is different
for global long-range connections via white matter, which show a slight increase in
average axon diameter with brain size \cite{wang}. Moreover these axons are myelinated, 
which enhances several-fold the speed of signal propagation in comparison to unmyelinated
intracortical axons. These empirical facts indicate that for global brain organization
the distance and delays are important constraints, and they were repeatedly used by
many researchers to model large-scale organization of brain connectivity 
\cite{cherniak,klyachko,kaiser,chen,cherniak2004,perez}. Some of these studies initially 
showed that certain parts of the nervous system in macaque monkey (prefrontal areas) 
and in the nematode {\it C. elegans} are optimized for wiring length 
\cite{cherniak,klyachko,perez}. However, more recent studies demonstrated that wire 
length was not fully minimized across macaque and cat visual areas \cite{cherniak2004}, 
and more importantly, global connections in the whole networks of macaque and {\it C. elegans} 
brains are far from being optimal for wire cost \cite{kaiser,chen}. The latter studies 
indicate that there might be other constraints on global brain organization, such as 
the requirement of short processing paths, which was proposed in the past 
\cite{karbowski2001}, or some other combinations 
\cite{wen2009,budd2010,budd2012,bullmore,karbowski2014}.

The important issue is how to relate local organization in the cortex to global
cortico-cortical connectivity. This is a challenging task, not only conceptually,
but also from a methodological point of view. The approach presented here for the
description of local networks relies on analytical optimization of Lagrange functions,
and it differs considerably from the approaches used for studying patterns of long-range
connections, which use mainly numerical algorithms of graph theory \cite{bullmore,kaiser,chen}.
It is not clear how to combine the two approaches within a single mathematical framework.
However, despite these technical difficulties, there is at least one quantity that can
relate local and global organizations. This quantity is associated with the fraction of
axons in the cortex $x$, which is composed of two contributions: local intracortical axons
and endings of the long-range axons (via white matter). Empirical data indicate that the
latter component is substantial \cite{stepanyants}, and it seems to be possible
to find a mathematical formula relating the two contributions. Since the long-range
part of $x$ should be somehow correlated with the volume of axons in white matter, 
it may be feasible to make a connection between variables operating locally with
those operating globally. In particular, one might try to derive from ``first principles''
the scaling relation between volumes of gray and white matters.

\newpage

\noindent
{\bf Generalization of the optimality models.} \\
The principle of spine economy maximization (as well as other principles) was defined
locally in this paper. This means that spatial correlations between different cortical
components were neglected. One natural extension of this work is to include spatial
dependence in the fitness functions (Eqs. 1, 21, and 24), by considering more local
circuits with slightly different properties that are coupled together. Such an approach 
would allow us, in principle, to model spatial plasticity effects and competition for
space in the cortex. For example, it is known that learning modifies the structure
of spines \cite{kasai}, and there are some indications that it also alters dendritic
and axonal processes \cite{chklovskii2004}. The interesting question here is how
these two types of modifications are related to one another, and how they influence
neighboring circuits.

Another possible extension is to include explicitly the temporal aspect in the
equations. The current approach, without time, describes a mature ``average'' brain.
Time dependence of cortical composition would allow us to model the effects
associated with brain development. In particular, there are data on synaptic
density development in different parts of the cortex across species (references
in \cite{karbowski2012}). Many of these dependencies show that synaptic density
acquires a maximum at some early developmental stage, and then it decays to
adult (stable) values. It would be interesting to see if spine economy rule
combined with time can generate such non-monotonic dependencies. The temporal
aspect in the equations could formally be included in analogy with a Hamiltonian
approach known from classical mechanics \cite{landau}, i.e. the fitness function 
$F_{s}$ could serve as a Hamiltonian of the local circuit.

\newpage

\noindent
{\bf Conclusions.} \\
This study shows that hierarchical composition of local cortical circuits can be
best explained by two different design principles. One is associated with a new
proposition called here spine economical maximization, and another with neural wire
volume minimization, and both give similar the best optimal solutions.
In contrast, other principles related to wire minimization such as: wire length,
wire surface area, or conduction delays minimizations do not yield reasonable 
results. Only when combined with spine economy rule, these other notions can
equally well be fitted to the data under some conditions (spine economy contribution
must dominate in the mixing of concepts in the meta-fitness function and/or spine size 
distribution must have a long tail). These results imply that for the efficiency 
of local circuits (i) wire volume may be more basic variable than wire length 
or temporal delays, (ii) spine economy principle may be an important concept, and
(iii) we should pay more attention to spines, especially in a broader context of 
brain evolutionary design.

\newpage

{\large \bf METHODS}

\vspace{0.3cm}

\noindent
{\large \bf Data gathering and analysis for cortical composition.}

\vspace{0.2cm}

\noindent
{\bf The ethics statement does not apply to this study}. \\
Experimental data in Table 1 for cortical composition come from different sources. 
They are either directly taken from a source or calculated based on other related 
neuroanatomical data.

\noindent
{\bf Mouse data.} Data from \cite{braitenberg}, except for capillaries - data 
from \cite{tsai}.  

\noindent
{\bf Rat data.} Data from \cite{mishchenko}, except for capillaries - data 
from \cite{bar}. 

\noindent
{\bf Rabbit data.} Fractional volumes are arithmetic means of the values for cortical 
spaces between and within dendrite bundles in the visual cortex of layers 2 and 3 
\cite{schmolke}.  

\noindent 
{\bf Cat data.} Density of axon length and dendrite length near the layer 3/4 of 
visual cortex was estimated as respectively $3.93\pm 0.8$ $\mu$m/$\mu$m$^{3}$ and 
$0.39 \pm 0.08$ $\mu$m/$\mu$m$^{3}$ \cite{stepanyants}. The fractional volumes of 
axons and dendrites were obtained by assuming axon and dendrite diameters as 
respectively 0.3 $\mu$m \cite{braitenberg} and 1.0 $\mu$m \cite{mainen}. Astrocyte 
data come from \cite{williams}, and capillary data from \cite{pawlik}.    

\noindent
{\bf Macaque monkey data.} The fractional volume of dendrites was estimated in 
prefrontal cortex as $0.33 \pm 0.19$, based on the formula: 
$(\pi/4)\rho_{n}l_{d}d^{2}$, where the neuron density $\rho_{n}= (1.0\pm 0.2)10^{5}$ 
mm$^{-3}$ \cite{christensen}, average total dendrite length per neuron 
$l_{d}= 3478 \pm 99$ $\mu$m \cite{hao}, 
and average dendrite diameter $d= 1.1 \pm 0.2$ $\mu$m \cite{escobar,medalla}.
The volume fraction of spines was estimated as a product of average spine head 
volume $0.15\pm 0.01$ $\mu$m$^{3}$ (probably an underestimate for a whole 
spine volume) and spine density $0.30\pm 0.03$ $\mu$m$^{-3}$ in prefrontal cortex 
\cite{dumitriu}. Capillary fraction data come from visual cortex \cite{weber}.  

\noindent
{\bf Human data.} Average spine volume (cingulate cortex) is $0.35 \pm 0.02$ 
$\mu$m$^{3}$ \cite{benavides} and density of asymmetric synapses (temporal cortex), 
presumably spines, is $(4.23 \pm 2.59)10^{-1}$ $\mu$m$^{-3}$ \cite{alonso}. 
Average fraction of spine volume was estimated as the product of these two parameters. 
The ratio of cortical volumes taken by dendrites and by spines is 2.39, which comes
from dividing total volumes of basal and apical dendrites (424.5 $\mu$m$^{3}$) and
spines (177.3 $\mu$m$^{3}$) per neuron in cingulate cortex of 40 years old human
\cite{benavides}. The volume fraction of dendrites was estimated as a product of 
2.39 and the volume fraction of spines. Astrocyte and capillary fractions data come 
from parietal cortex \cite{virgintino}.

\vspace{0.95cm}

\noindent{\large \bf Theoretical modeling.}

\vspace{0.2cm}

To find the optimal structural layout of the mature cerebral cortex we slightly 
simplify the analysis and consider its five major components that seem to be 
functionally important: axons, dendrites, spines, glia/astrocytes, and capillaries. 
In the considerations below we rely on a concept of geometric probability, which 
relates fractional volumes of cortical components with average probabilities of 
their encountering. In this mean-field approach, the details of neuronal or glial 
arborizations are not important.

\vspace{0.35cm}

\noindent
{\bf Model of spine fractional volume.}  \\
In a mature brain axons and dendrites are much more structurally stable than
synapses (spines), which can change volume or even disappear relatively fast 
\cite{bonhoeffer,holtmaat,statman,meyer}. Consequently, it is assumed that axonal and 
dendritic fractions form two primary independent variables that set to a large 
degree the cortical layout. A third, independent variable is an average spine 
volume. Spine size is indirectly related to the amount of metabolic energy it 
uses, through Na/K-ATPase pumps located on spine membrane 
\cite{attwell2001}. Bigger spines with larger surface area require 
more energy for pumping out Na$^{+}$ ions and maintaining their concentration 
gradient than smaller spines. There is some evidence that dendritic spines are 
the major energy users in the cortex \cite{alle,harris,karbowski2012}, and thus, 
their energy or alternatively their volume, seems to be an important variable. In 
this study, we focus explicitly on spine volume as an independent variable instead 
of spine energy, since there exist data on the distribution of spine sizes 
\cite{benavides,arellano,loewenstein} or spine EPSP 
\cite{song}, and no data on spine energy distribution.

A synaptic connection between excitatory neurons, which are majority in the cortex
\cite{braitenberg}, can potentially be generated in that cortical region where 
axonal and dendritic trees spatially overlap \cite{peters}. However, a physical 
vicinity of axons and dendrites may not be sufficient for the appearance of a spine 
in this location \cite{shepherd}, which suggests an additional factor of possibly 
stochastic nature involved \cite{loewenstein}.
It is assumed here that this factor is associated with energy. Specifically, we
require that a metabolic energy allocated to a spine must be above a certain threshold
to form a stable spine (or equivalently that potential spine volume must be large 
enough). This requirement relates to the empirical fact that physiological processes 
need a minimal amount of energy to be activated \cite{baron}, which seems to apply 
to dendritic spines, since they disappear during prolonged severe ischemia \cite{zhang} 
and during excessive cooling of the tissue \cite{kirov}. Thus, the average 
probability of finding a spine at some location is equal to the product of
two average probabilities: that axons and dendrites are present there, and that 
a potential spine is larger than a certain threshold. On the other hand, based on 
the concept of geometric probability, the probability of finding a spine in the 
cortex is approximately equal to the fractional cortical volume occupied 
by spines, which is denoted as $s$. Mathematically, this means:

\begin{eqnarray}
s= P xy,  
\end{eqnarray}\\
where the parameters $x$ and $y$ denote fractional volumes of axons and 
 dendrites in the cortex or probabilities of their occurrence (i.e. $0 < x, y < 1$).
The product $xy$ is the average probability that both axon and dendrite are
present in a small cortical space. The symbol $P$ is the conditional probability 
that a spine has volume $u$ that is greater than the threshold $\theta$. Formally, 
this probability is defined as

\begin{eqnarray}
P= \int_{\theta}^{\infty} H(u) du, 
\end{eqnarray}\\
where $H(u)$ is the distribution (density of probability) of spine volumes 
$u$. One can view $P$ as a conditional probability of spine formation.
We consider five different types of volume distribution functions $H(u)$, 
for which we obtain different forms of the probability $P(\overline{u})$ as 
a function of average spine volume $\overline{u}$
(see below).

\newpage

\noindent
{\bf Distributions of spine sizes.}  \\
Empirical data show that spines can have widely different sizes, from very small
$\sim 0.01$ $\mu$m$^{3}$ to quite large  $\sim 1.3$ $\mu$m$^{3}$ 
\cite{benavides,yasumatsu}. The distribution of their sizes has been fitted by 
two distinct functions with different asymptotic properties, either by gamma 
function with short-tail \cite{benavides}, or by log-normal with heavy-tail 
\cite{loewenstein}. This suggests that there could also be other distributions, 
which are statistically indistinguishable from the above,
that would fit the data equally well. For this reason and for a larger generality 
we consider five different distributions $H(u)$ of spine volumes: three with 
short-tail and two with heavy-tail. For each distribution we provide explicit 
forms of the conditional probability $P$ in terms of the average spine volume 
$\overline{u}$, which is defined as

\begin{eqnarray}
\overline{u}= \int_{0}^{\infty} H(u) u du.
\end{eqnarray}\\

\vspace{0.15cm}

\noindent \underline{Exponential distribution.} \\
This type of spine volume distribution has the form:

\begin{eqnarray}
H(u)= \alpha e^{-\alpha u}
\end{eqnarray}\\
for $u \ge 0$, where $\alpha$ is some positive constant. The average 
spine volume is $\overline{u}= 1/\alpha$. The conditional probability $P$ 
(defined in Eq. 3) that a spine is greater than the threshold $\theta$ is 

\begin{eqnarray}
P(\overline{u})= e^{-\theta/\overline{u}},
\end{eqnarray}\\
i.e. it can be expressed as a function of $\overline{u}$. The latter feature 
applies to all size distributions considered in this paper (see below).
For $\overline{u}/\theta \ll 1$ the probability $P(\overline{u}) \ll 1$, 
whereas for $\overline{u}/\theta \gg 1$ we have  $P(\overline{u}) \approx 1$.
 
\vspace{0.15cm}

\noindent \underline{Gamma distribution.} \\
The distribution of spine volume $H$ is: 

\begin{eqnarray}
H(u)= \frac{\alpha^{n+1}}{n!} u^{n} e^{-\alpha u}
\end{eqnarray}\\
for $u \ge 0$, where $\alpha$ is some positive constant. The average 
spine volume is  $\overline{u}= (n+1)/\alpha$. 
The probability 
$P= \Gamma\left(n+1,\frac{(n+1)\theta}{\overline{u}}\right)$, where 
$\Gamma$ is the standard Gamma function. In the paper we consider two special 
cases, with $n=1$ and $n=2$, for which the probability $P$ takes the following 
forms:

\begin{eqnarray}
P(\overline{u})= \left(1 + 2\frac{\theta}{\overline{u}}\right) 
e^{-2\theta/\overline{u}}
\end{eqnarray}\\
for $n=1$, and

\begin{eqnarray}
P(\overline{u})= \left[1 + 3\frac{\theta}{\overline{u}} + 
\frac{9}{2}\left(\frac{\theta}{\overline{u}}\right)^{2}\right] 
e^{-3\theta/\overline{u}}
\end{eqnarray}\\
for $n=2$.

\vspace{0.15cm}

\noindent \underline{Rayleigh distribution.} \\
The distribution of spine volume is given by

\begin{eqnarray}
H(u)= \frac{u}{\sigma^{2}} e^{-u^{2}/(2\sigma^{2})}
\end{eqnarray}\\
for $u \ge 0$. The average spine volume is  
$\overline{u}= \sqrt{\frac{\pi}{2}}\sigma$. The probability $P$ that 
spine is larger than the threshold $\theta$ is

\begin{eqnarray}
P(\overline{u})= e^{-\frac{\pi}{4}(\theta/\overline{u})^{2}}.
\end{eqnarray}\\

\vspace{0.15cm}

\noindent \underline{Log-logistic distribution.} \\
This type of distribution has a heavy tail and is represented by

\begin{eqnarray}
H(u)= \frac{\beta}{\alpha} \frac{(u/\alpha)^{\beta-1}}
{\left[1 + (u/\alpha)^{\beta}\right]^{2}}
\end{eqnarray}\\
for $u \ge 0$, where $\alpha$ is a positive constant and $\beta > 1$. 
Note that $H(u)$ decays as a power law for asymptotically large $u$.
The average spine volume is 
$\overline{u}= \alpha\pi/(\beta\sin(\pi/\beta))$. The probability
$P$ in terms of $\overline{u}$ is given by

\begin{eqnarray}
P(\overline{u})= \frac{\overline{u}^{\beta}}
{\overline{u}^{\beta} + \tilde{\theta}^{\beta} },
\end{eqnarray}\\
where $\tilde{\theta}$ is the renormalized threshold, i.e.
$\tilde{\theta}= \theta \frac{(\pi/\beta)}{\sin(\pi/\beta)}$.
Eq. (13) is known as Hill equation and is often used in biochemistry when
there are cooperative phenomena between different molecules. Note that
for sufficiently large exponent $\beta$, the probability $P \ll 1$ if
$\overline{u} < \tilde{\theta}$, and $P\sim 1$ if
$\overline{u} > \tilde{\theta}$.

\vspace{0.15cm}

\noindent \underline{Log-normal distribution.} \\
The distribution of spine volume in this case also has a heavy tail, and 
it is given by

\begin{eqnarray}
H(u)= \frac{1}{\sqrt{2\pi}\sigma u} 
\exp\left[-\frac{(\ln u -\mu)^{2}}{2\sigma^{2}}\right],
\end{eqnarray}\\
where $\mu$ and $\sigma$ are some parameters ($\sigma > 0$). The average
spine volume $\overline{u}$ is $\overline{u}= \exp(\mu + \sigma^{2}/2)$.
The probability that a spine has larger volume than the threshold $\theta$
is

\begin{eqnarray}
P(\overline{u})= \frac{1}{2}\left[1 - 
\mbox{erf}\left(\frac{\ln(\theta/\overline{u}) + \sigma^{2}/2}{\sqrt{2}\sigma}\right)
\right],
\end{eqnarray}\\
where erf(...) is the standard error function. Note that for $\sigma \ll 1$,
we have $P \sim 1$ if $\overline{u}/\theta \gg 1$ and $P \ll 1$ if
$\overline{u}/\theta \ll 1$.

\vspace{0.35cm}

\noindent
{\bf Model of glia and capillary fractional volumes.}  \\
The model of glia and capillary fractions presented below relies on empirical 
evidence that cortical neurons with their synapses are spatially coupled to 
glia (astrocytes) and microvasculature. Specifically, it was shown that neuron
density for mouse cortex and macaque monkey visual cortex correlate with vascular 
length density \cite{tsai,weber}. The changes in astrocyte density are to some 
extent related to changes in capillary density across layers 
of mouse somatosensory cortex \cite{mccaslin} (see below). Moreover, 
developmental data for cat visual cortex show a strong spatio-temporal coupling 
between capillary length density and synaptic density \cite{tieman}.

Extended branching processes of astrocytes resemble dendritic trees of neurons
\cite{oberheim2006}. The endings of these processes physically connect with spines 
by wrapping around their surface to provide spines with metabolic substrates
and glutamate from capillaries, and to remove waste products \cite{oberheim2006}. 
The transport of metabolites to and from spines along astrocytes should be effective, 
i.e. sufficiently fast and the least energy consuming, otherwise spines and thus 
local neural circuits would not get enough energy on time and as a result would 
underperform their functions. (Although glia use much less energy than neurons 
\cite{attwell2001}, they nevertheless should minimize their metabolic needs for 
the overall brain efficiency). This suggests that the total length of astrocyte 
processes should be as small as possible. Consequently, we can treat an astrocyte 
as a minimal spanning tree along its target points, i.e., spines. It can be shown
mathematically that in a minimal tree formalism, the total length of the tree 
connecting $n$ target points or branch points scales asymptotically as $n^{2/3}$ 
\cite{beardwood,steele}. The important point is that this result applies universally 
to any transportation network that works efficiently. Additionally, because of the 
physical units consistency, the total length of the tree should scale with enclosing 
volume $V$ as $V^{1/3}$. Thus, the total length $L$ of processes of a single astrocyte 
connecting $N_{s}$ spines should be minimized when 

\begin{eqnarray}
L= b N_{s}^{2/3} V^{1/3},  
\end{eqnarray}\\
where $b$ is some constant. It can be shown theoretically under general conditions
that $b= (3/(4\pi))^{1/3}$ \cite{cuntz}. This value was also found empirically
for neural dendritic trees \cite{cuntz}, and because of the structural similarity 
between branching patterns of astrocytes and dendrites, this particular value
of $b$ is also adopted here.  The parameter $V$ in Eq. (16) is the cortical volume 
enclosing a single astrocyte and $N_{s}$ spines. This ``domain volume'' is defined by 
three-dimensional boundaries of an astrocyte, and it is much bigger than the actual 
volume of an astrocyte because it includes also other cortical components (spines, 
dendrites, axons, etc) contained within these boundaries. 
The critical feature of astrocyte domains is that neighboring astrocytes 
essentially do not overlap, which means that each domain contains only one astrocyte 
cell that influences synaptic spines only from that particular domain. This
property of astrocytes spatial arrangement is called domain organization
(\cite{oberheim2006,ogata}).

The main contribution to astrocyte volume comes from astrocyte free processes
(astrocyte soma and astrocyte perivascular sheath constitute only 28$\%$ and
7$\%$ of the total astrocyte volume; \cite{virgintino}).
Assuming a cylindrical geometry for these extended processes, the total
volume of an astrocyte $V_{as}$ can be  approximated as
$V_{as}= (\pi/4)L d_{as}^{2}$, where $d_{as}$ is the average diameter of all
free processes. The value of $d_{as}$ can be estimated based on data for
volume ($V_{pr}= 350$ $\mu$m$^{3}$) and surface area ($S_{pr}= 1650$ $\mu$m$^{2}$)
of astrocyte processes (without soma) in the cat sensorimotor cortex
\cite{williams}. Using a familiar formula $d_{as} = 4V_{pr}/S_{pr}$,
we obtain $d_{as}= 0.85$ $\mu$m. One can expect that the value of $d_{as}$ only 
very weakly depends on brain size, as it is the case for diameters of the thickest 
processes, which are 2.2 $\mu$m for mouse, and 2.9 $\mu$m for human \cite{oberheim2009},
despite four orders of magnitude difference in brain volumes of these mammals
(e.g. \cite{karbowski2007}). For that reason, the value $d_{as}= 0.85$ $\mu$m is 
kept constant for all computations performed in this study, and it is the only 
parameter that is fixed in the model.

Because of the astrocyte domains segregation, the volume fraction of astrocytes 
can be defined as $g= V_{as}/V$, which combined with Eq. (16) yields

\begin{eqnarray}
g= a \rho_{s}^{2/3}= \frac{a s^{2/3}}{\overline{u}^{2/3}},  
\end{eqnarray}\\
where $a= (\pi/4)b d_{as}^{2}= 0.352$ $\mu$m$^{2}$, spine density 
$\rho_{s}= N_{s}/V$, and in the last equality we used the fact that
density $\rho_{s}$, spine proportion $s$, and average spine volume
$\overline{u}$ are related by

\begin{eqnarray}
s= \rho_{s}\overline{u}.
\end{eqnarray}

The spatial separation between capillaries and spines, and between capillaries 
and astrocytes is relatively small, which presumably ensures a high efficiency 
of energy delivery. For example, a typical distance 
between capillaries and spines in mouse cortex is roughly 13 $\mu$m \cite{zhang}. 
Many astrocyte processes are either in the vicinity or directly touch capillaries 
\cite{oberheim2006}. Even astrocyte somata is close to microvasculature, with mean 
spacing between them $6-10$ $\mu$m in mouse somatosensory cortex, which at some 
locations can be down to $\sim 1$ $\mu$m \cite{mccaslin}. 
For a comparison, an intercapillary 
distance is generally much larger: $32-43$ $\mu$m for mouse cortex \cite{boero}, 
and 58 $\mu$m for human cortex \cite{meier}.
These data suggest that capillaries cluster in those cortical places where there 
are high densities of both astrocytes and spines. 
Mathematically, this means that the probability of finding a capillary at some
location (or equivalently, fraction of capillary volume $c$) is proportional to 
probability of finding both an astrocyte (equal to volume fraction $g$) and 
a spine (equal to volume fraction $s$). The simplest form of such a probabilistic 
relationship is their product, i.e. 

\begin{eqnarray}
c= g s. 
\end{eqnarray}\\
There is some additional empirical support for that ``product formula'' based
on a laminar distribution of microvasculature and synapses both for rodents
and for primates. For mouse somatosensory cortex, capillary fraction and astrocyte 
density correlate across cortical layers \cite{mccaslin}, but their relationship 
is clearly nonlinear, as local peaks in these two variables often do not exactly match
(compare Fig. 1 in \cite{mccaslin}). It seems that inclusion of spine density 
in that relationship should improve the correlation with changes in capillary
fraction variability. Specifically, capillary fraction exhibits a peak in cortical
somatosensory layer 1, which is however absent in the corresponding astrocyte density 
\cite{mccaslin}, but it correlates well with the fact that the density of 
asymmetric synapses (mostly spines) is the largest in the layer 1 of mouse 
somatosensory cortex \cite{defelipe}.

For primate visual cortex, capillary length density is the largest in the middle
layers 4 and 2/3, and the smallest in layers 1 and 5/6 \cite{weber,bell}, 
which is similar to the laminar distribution of synaptic density, although with 
some fluctuations \cite{okusky,huttenlocher}. 
Taken together, these data suggest that microvasculature is correlated with 
both synapses and astrocytes in the cortical gray matter.

Capillary fraction $c$ in Eq. (19) can be expressed in terms of spine parameters 
using Eq. (17) for $g$, with the result

\begin{eqnarray}
c= \frac{a s^{5/3}}{\overline{u}^{2/3}}.  
\end{eqnarray}\\
Note, that because $s \sim g^{3/2}$, we have equivalently that $c \sim g^{5/2}$,
which indicates a strong nonlinear dependence between capillary and astrocyte 
volume fractions, and suggests that generally one can expect $c/g \ll 1$.

\vspace{0.45cm}

\noindent
{\bf The fitness functions.}  \\
We consider three different classes of fitness functions. The first corresponds
to the principle of neural ``wire minimization'' and the second to the proposed here 
``spine economical maximization''. The third class is a combination of the first
two.

\vspace{0.15cm}

\noindent \underline{Wire minimization principle}. \\
The most general form of the fitness function $F_{w}$ (benefit-cost or Lagrange function)
for wire minimization takes the form:

\begin{eqnarray}
F_{w}= \frac{rx + y}{\overline{u}^{\gamma_{1}}} + \lambda_{1}\left(x + y + s + g + c - 1 \right),
\end{eqnarray}\\
where the exponent $\gamma_{1} > 0$ corresponds to specific characteristics of neuronal wire
one wants to minimize (wire length, its surface area, its volume
or conduction delays; see below).
The free parameter $r$ is a measure of the asymmetry between axons and dendrites.
The parameter $\lambda_{1}$ is the Lagrange multiplier associated with the volume 
normalization constraint: $x + y + s + g + c = 1$, i.e., fractional volumes of 
all considered cortical components must sum up to unity. The fitness function 
$F_{w}(x,y,\overline{u})$ is the function of the three independent variables $x, y$, 
and $\overline{u}$, because the fractions $s$, $g$, and $c$ depend on these 
variables (see Eqs. (2), (17), and (20)). 

The form of the benefit-cost function in Eq. (21) can be justified by
taking neural wire length minimization as an example. The other cases can be analyzed 
analogously. The total axonal volume in the cortex is $V_{a}= (\pi/4)L_{a}d_{a}^{2}$, 
where $L_{a}$ is the total axon length in the cortex and $d_{a}$ is the average axon
diameter. Similarly, for the total dendrite volume we have $V_{d}= (\pi/4)L_{d}d_{d}^{2}$, 
where $L_{d}$ is the total dendrite length and $d_{d}$ is its diameter.
Now, consider the fitness function $F_{w0}$ as a density of combined lengths of 
axons and dendrites with some proportion coefficient $r_{0}$: 

\begin{eqnarray}
F_{w0} \sim \frac{r_{0}L_{a} + L_{d}}{V}, 
\end{eqnarray}\\
where $V$ is the volume of cortical gray matter. If we represent $L_{a}$ and
$L_{d}$ in terms of $V_{a}$ and $V_{d}$, and denote the fractional volumes of
axons and dendrites respectively as $x= V_{a}/V$ and $y= V_{d}/V$, then we
obtain that 

\begin{eqnarray}
F_{w0} \sim \frac{(d_{d}/d_{a})^{2}r_{0}x + y}{d_{d}^{2}}. 
\end{eqnarray}\\
The empirical data indicate that average spine heads (postsynaptic density PSD) 
and dendrites have diameters of the same order of magnitude (fraction of micron),
which do not seem to depend significantly on brain size \cite{karbowski2014}.
This suggests that these two diameters can be mutually coupled, which implies that 
one can assume that $d_{d}^{2}\sim \overline{u}^{\gamma_{1}}$, where the exponent 
$\gamma_{1}\approx 2/3$ (the bulk of the spine has a spherical shape). 
Moreover, if we denote $r= r_{0}(d_{d}/d_{a})^{2}$, then we obtain Eq. (21) 
for the full (with the Lagrange multiplier term) fitness function.

A similar analysis performed for wire surface minimization and wire volume 
minimization yields the same formula but with different $\gamma_{1}$, respectively
1/3 and 0. For the case of conduction delays minimization, one can define a similar
fitness function to $F_{w0}$ in Eq. (22) with the substitutions 
$L_{a} \mapsto L_{a}/\sqrt{d_{a}}$ and $L_{d} \mapsto L_{d}/\sqrt{d_{d}}$, 
which approximately correspond to temporal delays along axons and dendrites, 
as the conduction velocity is proportional to a square root of unmyelinated wire 
diameter \cite{hursh}. Performing the analysis analogously, one obtains Eq. (21)
with $\gamma_{1}= 5/6$ and $r= r_{0}(d_{d}/d_{a})^{5/2}$.

\vspace{0.25cm}

\noindent \underline{Spine economical maximization principle}. \\
The most general form of the fitness function $F_{s}$ for spine proportion economical 
maximization takes the form:

\begin{eqnarray}
F_{s}= \frac{s}{\overline{u}^{\gamma_{2}}} + \lambda_{2}\left(x + y + s + g + c - 1 \right),
\end{eqnarray}\\
where the exponent $\gamma_{2} > 0$ characterizes the influence of spine size on the
maximization process ($\gamma_{2}$ here is generally numerically different than $\gamma_{1}$ 
in Eq. (21)), and the parameter $\lambda_{2}$ is the Lagrange multiplier as before. This 
proposition can be justified as follows. The analysis of experimental data on the 
developing cerebral cortex in several mammals indicate that the ratio of cerebral 
metabolic rate CMR (the rate of glucose consumption per cortical volume) and synaptic 
density $\rho_{s}$ is approximately conserved from birth until adulthood
for a given region of the cortex, despite large variabilities in CMR and $\rho_{s}$ 
\cite{karbowski2012}. This means that CMR/$\rho_{s}$ $\approx$ const. Synaptic density 
is comprised mostly of the density of spines (synapses in the cerebral cortex are in 
$80-90 \%$ excitatory, most of which are axo-spinal \cite{zecevic}, and this percentage 
does not depend on brain size; \cite{karbowski2014}), which is related to spine 
fractional volume $s$ via the formula $s=\rho_{s}\overline{u}$. 
This implies that the ratio CMR$\overline{u}/s$ is roughly
conserved during development. The product in the nominator can be identified as an 
average metabolic energy per spine, which should be proportional to spine volume at 
some power $\gamma_{2} \ge 0$, i.e. CMR$\overline{u} \sim \overline{u}^{\gamma_{2}}$, 
because spine energy is associated mainly with the activity of Na/K-ATP pumps located 
on spine surface area \cite{attwell2001,erecinska}. Thus, we obtain that the ratio 
$s/\overline{u}^{\gamma_{2}}$ should be developmentally approximately constant for a given 
cortical area, which indicates that this ratio is probably important for cortical 
functioning. Consequently, it is assumed here that there has been an evolutionary 
pressure on increasing the proportion of spines in the cortex that would consume the 
least energy or take the smallest size (spines are energy demanding 
\cite{alle,harris,karbowski2012}). These considerations lead to the maximization of 
the benefit-cost function given by Eq. (24).

\vspace{0.25cm}

\noindent \underline{Combined wire minimization and spine economical maximization as a meta 
principle}. \\
The simplest fitness function $F$ that generalizes both notions of wire minimization 
and spine maximization, and includes them simultaneously is a linear combination of their
corresponding primary fitness functions $F_{w}$ and $F_{s}$. If part of $F$ associated
with $F_{w}$ is to be minimized, and part of $F$ related to $F_{s}$ is to be maximized,
then we must include $F_{w}$ and $F_{s}$ with opposite signs. Therefore, we define
the meta fitness function $F$, which we want to minimize, as

\begin{eqnarray}
F= f F_{w} - (1-f)F_{s},
\end{eqnarray}\\
where $f$ is the parameter controlling relative contributions of wire minimization
and spine economy maximization Lagrangians, with the condition $0 \le f \le 1$.
As we gradually decrease $f$ from 1 to 0, then the meta fitness function $F$ changes
its character form dominated by wire minimization to dominated by spine economy 
maximization. The case $f= 1/2$ corresponds to a symmetric situation when both wire 
min and spine max rules are equally important. Note that minimization of $(-F_{s})$ 
is mathematically equivalent to maximization of $F_{s}$. 
Equation (1) in the Results section is obtained after substitution of Eqs. (21) 
and (24) for $F_{w}$ and $F_{s}$ in Eq. (25).

\vspace{0.7cm}

\noindent
{\bf Optimization of the fitness functions.}  \\
Optimal fractional volumes of axons $x$, dendrites $y$, spines $s$, glia/astrocytes $g$, 
capillaries $c$, and optimal average spine volume $\overline{u}$ are found by taking 
partial derivatives of $F$, i.e.,
$\partial F/\partial x = \partial F/\partial y = 
\partial F/\partial \overline{u} =  \partial F/\partial \lambda = 0$.
As a result we obtain a system of three basic equations (details are provided in Supp. 
Information S1 Text):

\begin{eqnarray}
(1-f)s(y-x) + f\overline{u}^{\gamma_{2}-\gamma_{1}}\left[xy(1-r)+(y-rx)
\left[s+\frac{g}{3}\left(2+5s\right)\right]\right]
= 0,
\end{eqnarray}\\

\begin{eqnarray}
\left[(1-f)s\overline{u}^{\gamma_{1}}+f\overline{u}^{\gamma_{2}}
\left(s+\frac{2}{3}g+\frac{5}{3}c\right)\right]
\frac{y\overline{u}}{P}\frac{\partial P}{\partial \overline{u}} =   \nonumber  \\
\left(y+s+\frac{2}{3}g+\frac{5}{3}c\right)\left[(1-f)\gamma_{2}s\overline{u}^{\gamma_{1}}
-f\gamma_{1}(y+rx)\overline{u}^{\gamma_{2}}\right]          \nonumber  \\
+ \frac{2}{3}\left(g+c\right)\left[fy\overline{u}^{\gamma_{2}}-(1-f)s\overline{u}^{\gamma_{1}}\right],
\end{eqnarray}\\
and
\begin{eqnarray}
x+y+s+g+c= 1.
\end{eqnarray}\\
In the case of pure wire minimization principle, i.e. when $f=1$, the above system reduces to
the following equations (with explicit dependencies of $s$, $g$, and $c$ on $x$, $y$, and $P$:

\begin{eqnarray}
(rx-y)\left[P + \frac{a}{3}\left(\frac{P^{2}}{xy\overline{u}^{2}}\right)^{1/3}
(2+5Pxy)\right] = 1- r ,
\end{eqnarray}\\

\begin{eqnarray}
 \frac{2}{3}aPx^{2/3}y(1+Pxy) - \gamma_{1}(rx+y)P^{1/3}\left[\overline{u}^{2/3}
y^{1/3}(1+Px) + \frac{a}{3}(Px)^{2/3}(2+5Pxy)\right]    \nonumber  \\
= \overline{u}\frac{\partial P}{\partial \overline{u}} 
\left[P^{1/3}\overline{u}^{2/3}xy^{4/3} + \frac{a}{3}x^{2/3}y(2+5Pxy)\right],
\end{eqnarray}\\
and

\begin{eqnarray}
x + y + Pxy +  \frac{a(Pxy)^{2/3}}{\overline{u}^{2/3}}
+  \frac{a(Pxy)^{5/3}}{\overline{u}^{2/3}} = 1.
\end{eqnarray}\\
In the case of pure spine economical maximization principle, i.e. when $f=0$, the system 
of Eq. (26-28) reduces to two equations:

\begin{eqnarray}
\overline{u}^{2/3} \frac{\partial P}{\partial \overline{u}} = 
\frac{P}{\overline{u}}
\left( \gamma_{2}\overline{u}^{2/3}(1 + Px)  +
\frac{a}{3}P^{2/3}x^{1/3}[2(\gamma_{2}-1) + (5\gamma_{2}-2)Px^{2}]\right)
\end{eqnarray}\\
and 

\begin{eqnarray}
2x + Px^{2} +  \frac{a(Px^{2})^{2/3}}{\overline{u}^{2/3}}
+  \frac{a(Px^{2})^{5/3}}{\overline{u}^{2/3}} = 1,
\end{eqnarray}\\
since in this case $x=y$.

The above equations are solved numerically for each distribution of spine volumes, 
using standard techniques \cite{press}. In the case of pure wire minimization,
the optimal solution corresponds to a local minimum of $F_{w}$, whereas for
pure spine economy maximization the optimal solution is associated with
a local maximum of $F_{s}$ (see S1 Text). For a mixed case with $0 < f < 1$,
the optimal solution is a local minimum of $F$ (S1 Text).

\vspace{1.2cm}

\noindent
{\bf Comparison of the theoretical results to the data.}  \\
The optimal theoretical fractional volumes were compared to the empirical fractional 
volumes in the cerebral cortex using two related measures of similarity: 
Euclidean distance (ED) and Mahalanobis distance (MD). The major difference between 
these two measures is in their treatment of variance in the data. ED measures 
a distance between theoretical points and mean values of data points ignoring their 
variance, whereas MD measures such a normalized distance including standard deviations 
of data points \cite{spencer}. Thus MD is more general than ED. The best similarity
with the data is achieved for minimal values of MD and ED.

ED distance between theoretical and experimental points is defined as:

\begin{eqnarray}
\mbox{ED}= \sqrt{ \sum_{i=1}^{5} (x_{i}-x_{ex,i})^{2} },
\end{eqnarray}\\
where $x_{i}$ and $x_{ex,i}$ are respectively theoretical and mean empirical values 
of fractional volumes of axons, dendrites, spines, glia/astrocytes, and capillaries
in the cortex (experimental data correspond to the next-to-last line in Table 1).

MD distance between theoretical and experimental points is defined as \cite{spencer}:

\begin{eqnarray}
\mbox{MD}= \sqrt{ \sum_{i=1}^{5} \left(\frac{x_{i}-x_{ex,i}}{sd_{ex,i}}\right)^{2} },
\end{eqnarray}\\
where $x_{i}$ and $x_{ex,i}$ are the same as in Eq. (34), and $sd_{ex,i}$ denotes
the standard deviation associated with each data point (the next-to-last line in 
Table 1). It is easy to see that MD defined above is simply a variance-normalized 
ED distance. In a special case when all standard deviations $sd_{ex,i}$ are equal 
to unity, MD reduces to ED.

\vspace{1.2cm}

\noindent
{\bf Supporting Information.}  \\
{\bf S1 Text.} This file contains Supporting Figure A and Supporting Tables A-E.
It also provides some details of the derivations, and proofs of minimum for $F_{w}$
and $F$, and proof of maximum for $F_{s}$. (PDF)


\newpage

\vspace{1.5cm}



\newpage

{\bf \large Figure Captions}

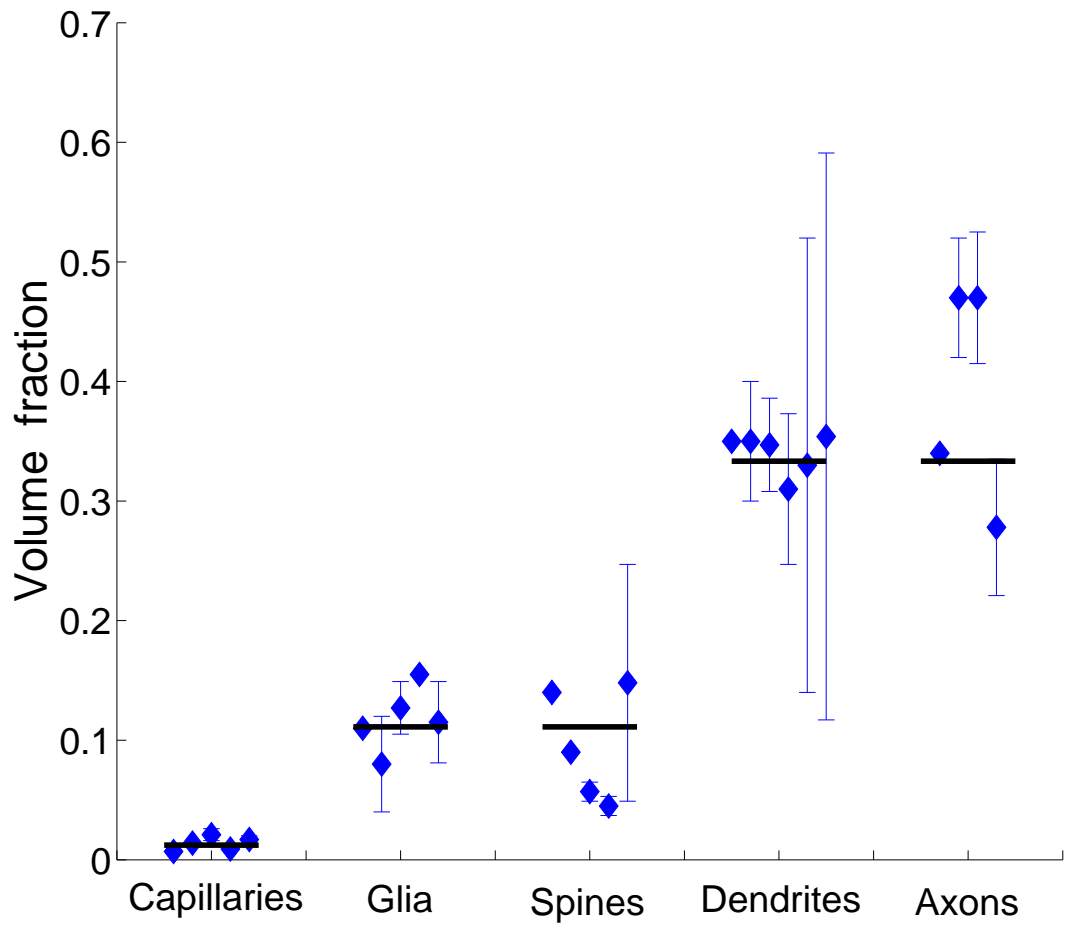
{\bf Fig. 1} \\
{\bf Hierarchical organization of the five major components in the gray matter of 
cerebral cortex.} 
Empirical data points for each component are denoted by
blue diamonds and correspond to different mammals (see Table 1). 
Solid black lines represent integer powers of the fraction 1/3. In particular, 
these fractions are: $(1/3)^{4}$ for capillaries, $(1/3)^{2}$ for glia and
spines, and $1/3$ for dendrites and axons. Note, that on average empirical 
fractional volumes across species approximately conform to this simple rule.

\vspace{0.3cm}

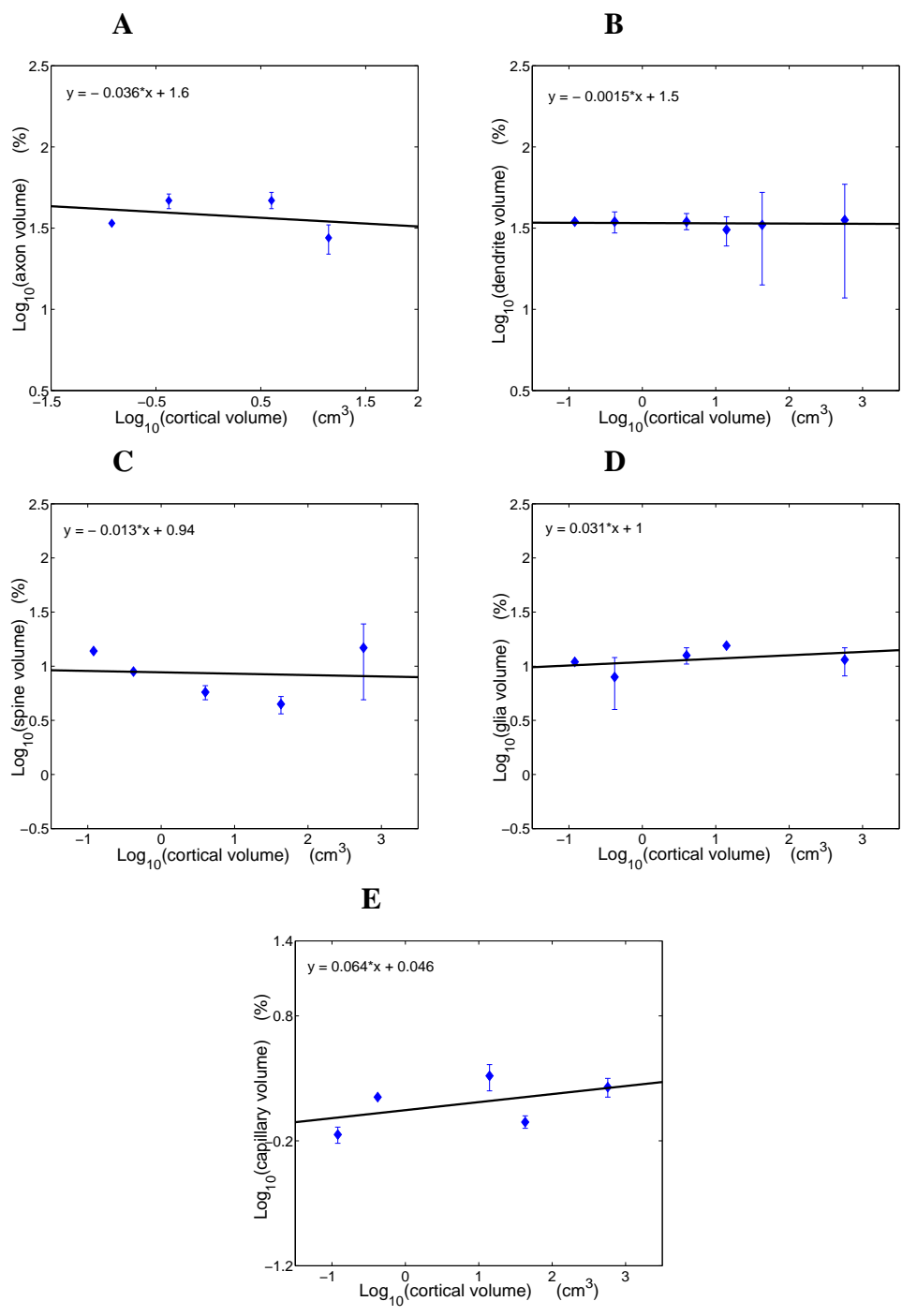
{\bf Fig. 2} \\
{\bf Scaling dependence of fractional volumes of the basic cortical components
on cortical gray matter volume.} (A) Axon fractional volume, (B) dendrite fractional
volume, (C) spine fractional volume, (D) glia fractional volume, and (E) capillary
fractional volume as functions of cortical volume in log-log coordinates.
Note a conservation trend across mammals.
Scaling plots were constructed based on data in Table 1. The following volumes 
of cortical gray matter (two hemispheres) were used:  mouse  0.12 cm$^{3}$ 
\cite{braitenberg}, rat 0.42 cm$^{3}$ \cite{houzel2010}, rabbit 4.0 cm$^{3}$ 
\cite{mayhew}, cat 14.0 cm$^{3}$ \cite{mayhew}, macaque monkey 42.9 cm$^{3}$ 
\cite{houzel2010}, human 571.8 cm$^{3}$ \cite{houzel2010}.

\vspace{0.3cm}

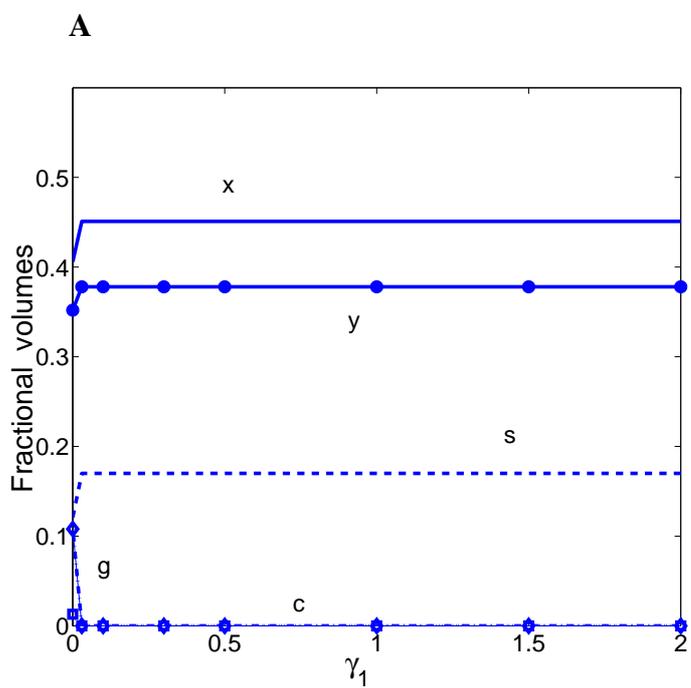
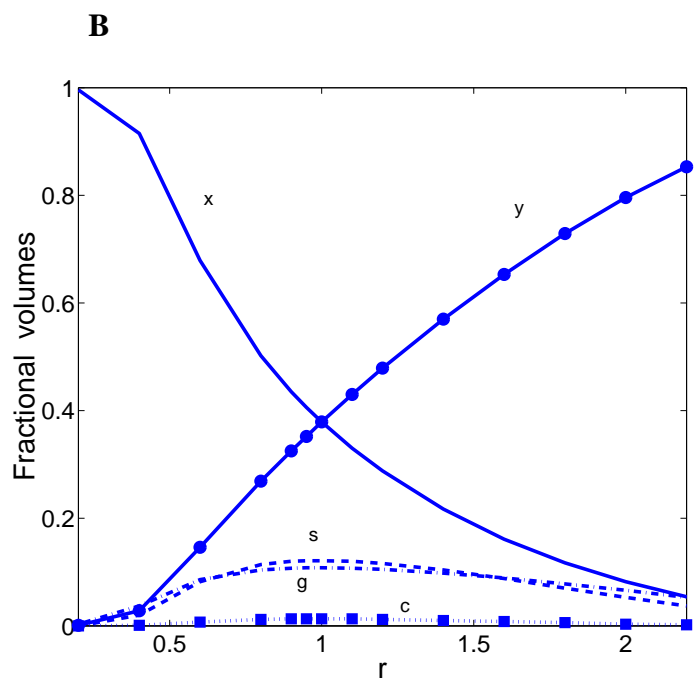
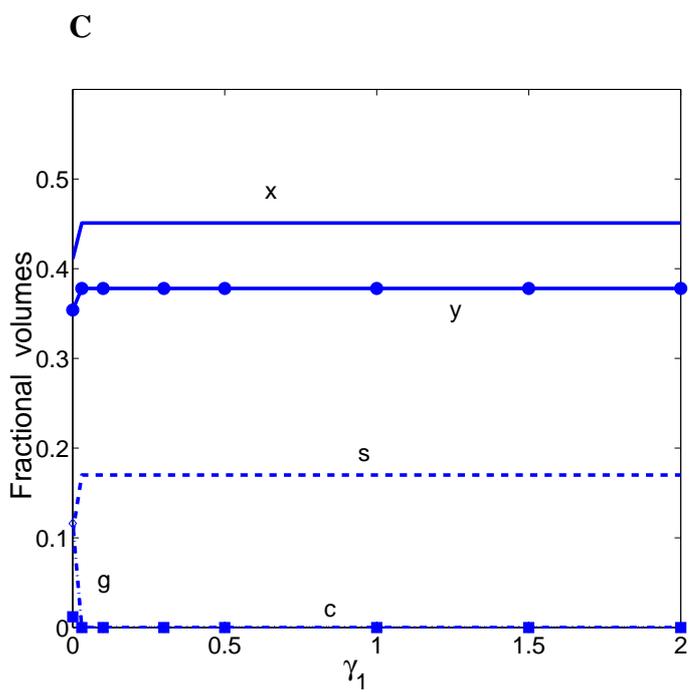
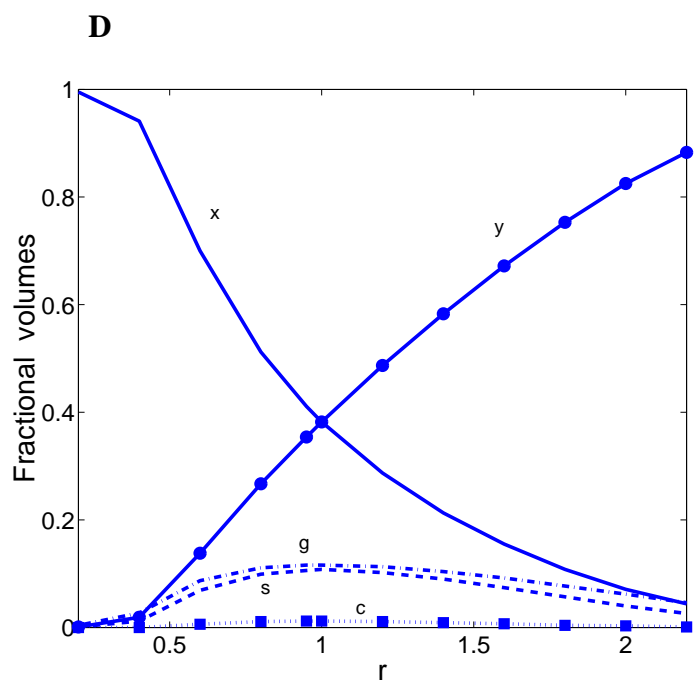
{\bf Fig. 3} \\
{\bf Optimal fractional volumes for neural ``wire minimization'' principle as 
functions of the exponent $\gamma_{1}$ and $r$.} 
The results for short-tailed Gamma (n=2) distribution are in (A) and (B), 
and for long-tailed Log-normal distribution in (C) and (D).
Note that for $\gamma_{1} > 0$ (panels A and C) all fractions are constant, 
in particular glia/astrocytes and capillaries vanish ($g= c= 0$).
Parameter values: for panels (A) and (C) $r=0.95$, while for panels
(B) and (D) $\gamma_{1}= 0$. Additionally, for log-normal $\sigma=0.3$.
For all panels $\theta= 0.321$ $\mu$m$^{3}$.

\vspace{0.3cm}

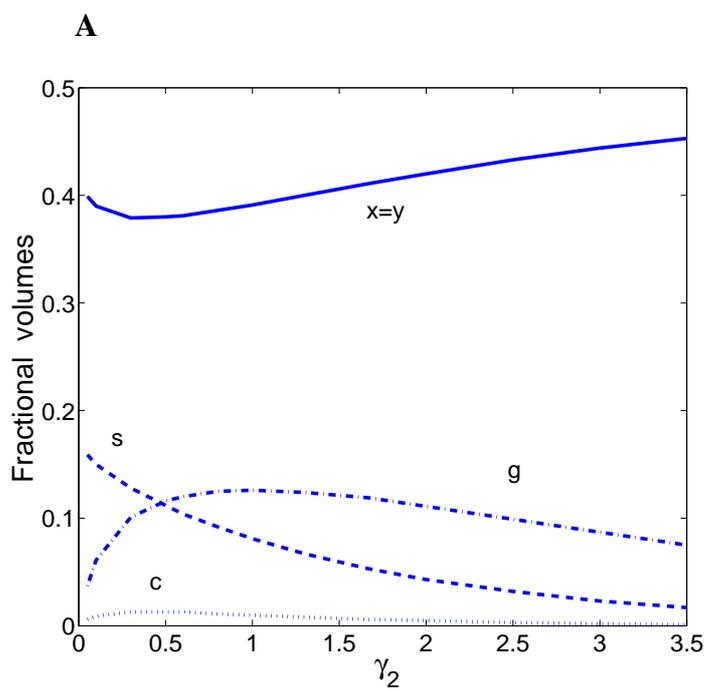
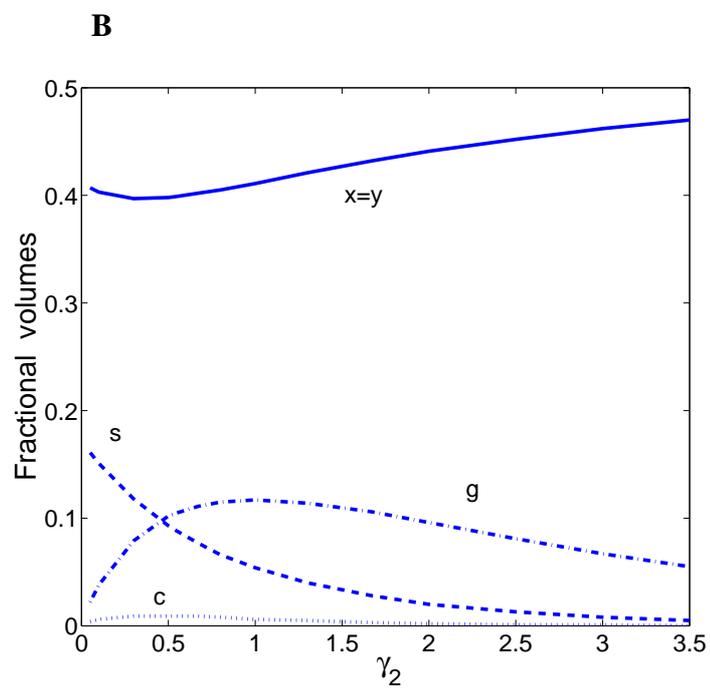
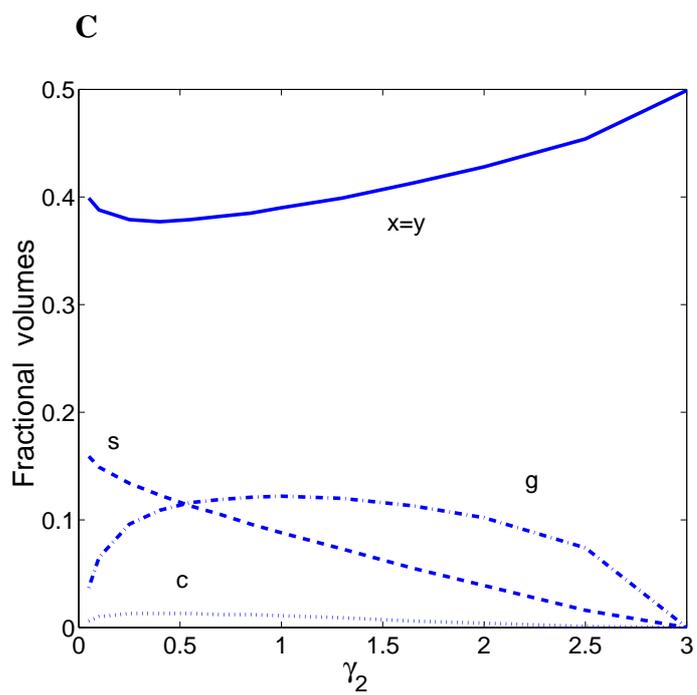
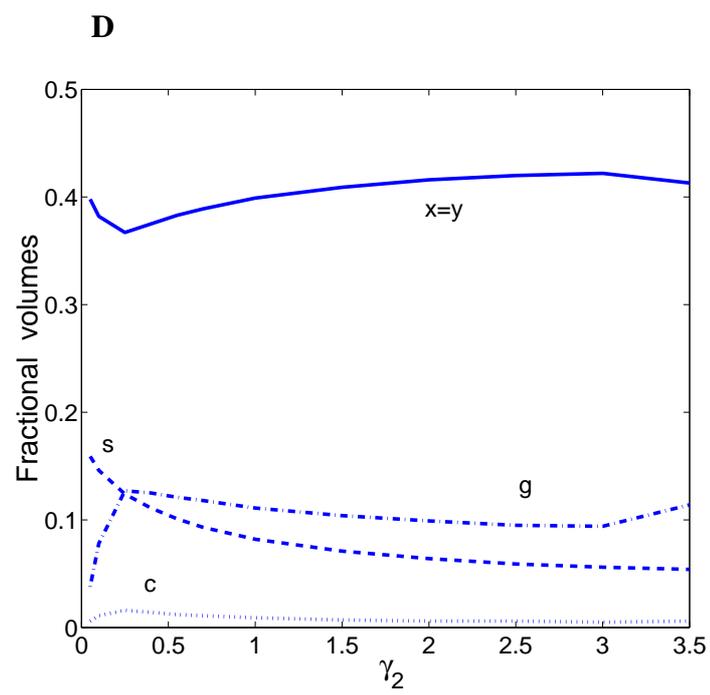
{\bf Fig. 4} \\
{\bf Optimal fractional volumes for ``spine economical maximization'' principle
as functions of the exponent $\gamma_{2}$.}
The results are qualitatively very similar for (A) Exponential distribution, 
(B) Gamma (n=2) distribution, (C) Log-logistic distribution ($\beta=3.0$), and 
(D) Log-normal distribution ($\sigma=0.25$). For all distributions 
$\theta= 0.321$ $\mu$m$^{3}$.

\vspace{0.3cm}

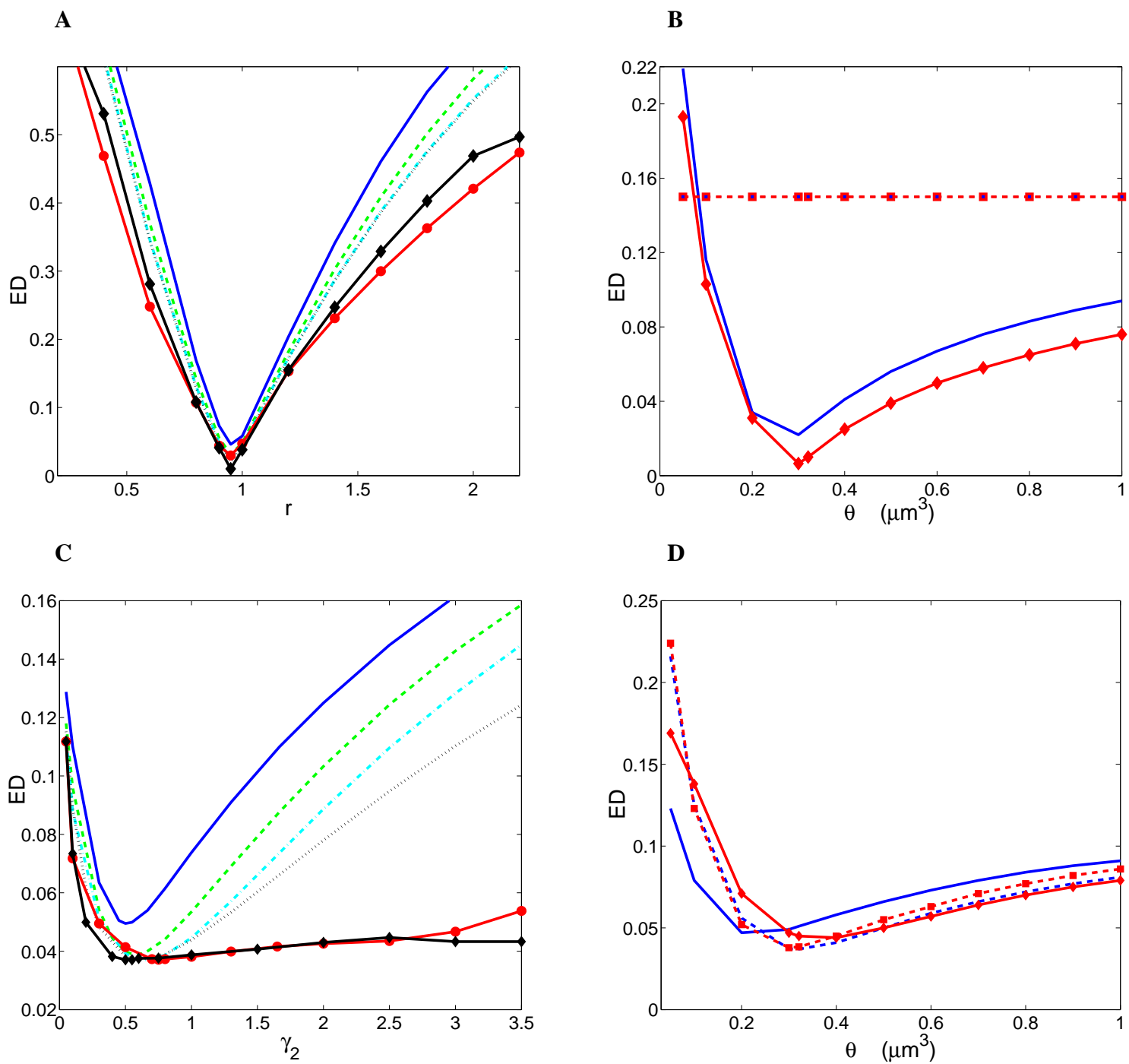
{\bf Fig. 5} \\
{\bf Euclidean distance (ED) between optimal theoretical results and empirical 
data as a function of $r$, threshold $\theta$, and $\gamma_{1}$, $\gamma_{2}$.} 
Panels (A) and (B) refer to ``wire minimization'', whereas panels (C) and
(D) correspond to ``spine economical maximization''.
(A) Dependence of ED on $r$ for all distributions of spine volumes for ``wire
volume minimization'', i.e. $\gamma_{1}=0$.
(B) ED as a function $\theta$ for $r=0.95$. Blue lines correspond to Gamma (n=2)
distribution and red lines (with diamonds and squares) to Log-normal 
($\gamma_{1}=0$ for solid lines and $\gamma_{1}=0.65$ for dashed lines). 
For Log-normal distribution $\sigma=0.3$.
Note that for $\gamma_{1} > 0$ ED is constant (dashed red line with blue squares),
i.e. ED= 0.15.
(C) Dependence of ED on $\gamma_{2}$ for all distributions of spine volumes.
(D) ED as a function $\theta$. Blue lines correspond to Gamma (n=2)
distribution and red lines (with diamonds and squares) to Log-normal 
($\gamma_{2}=0.3$ for solid lines and $\gamma_{2}=0.65$ for dashed lines). 
For Log-normal distribution $\sigma=0.25$.
In panels (A) and (C) $\theta=0.321$ $\mu$m$^{3}$, and the following labels
were used: Exponential distribution is shown as solid blue line, Gamma (n=1) 
and Gamma (n=2) are shown respectively as dashed green and dashdot cyan lines, 
Rayleigh distribution is represented by dotted black line, Log-logistic
is shown as solid red line with circles, and Log-normal as solid black
line with diamonds. The curves for Log-logistic and Log-normal correspond
to different values of the parameters, respectively, $\beta$ and $\sigma$ that
yield the minimal ED for a given $r$ or $\gamma_{2}$.

\vspace{0.3cm}

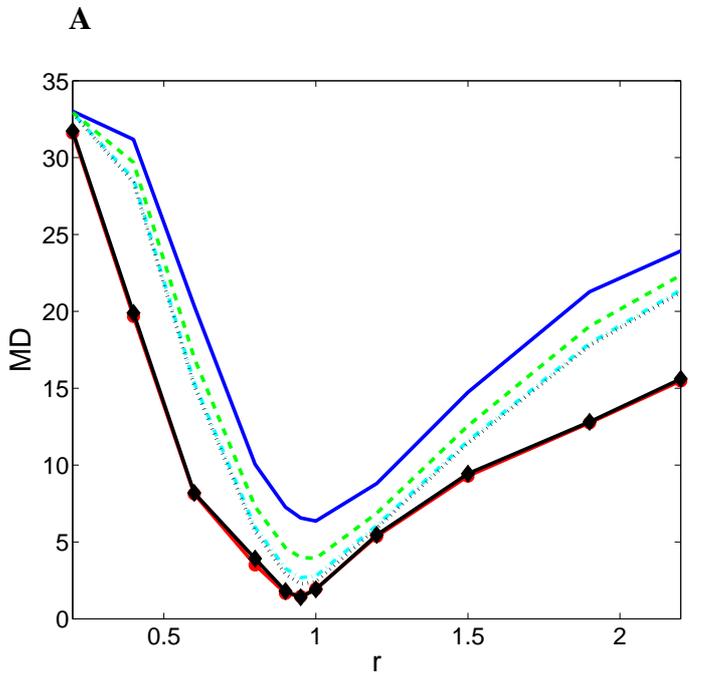
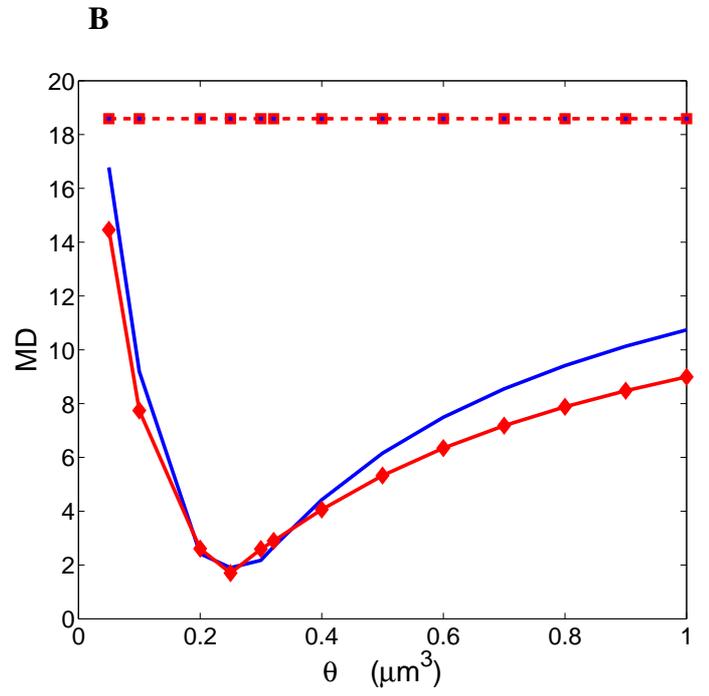
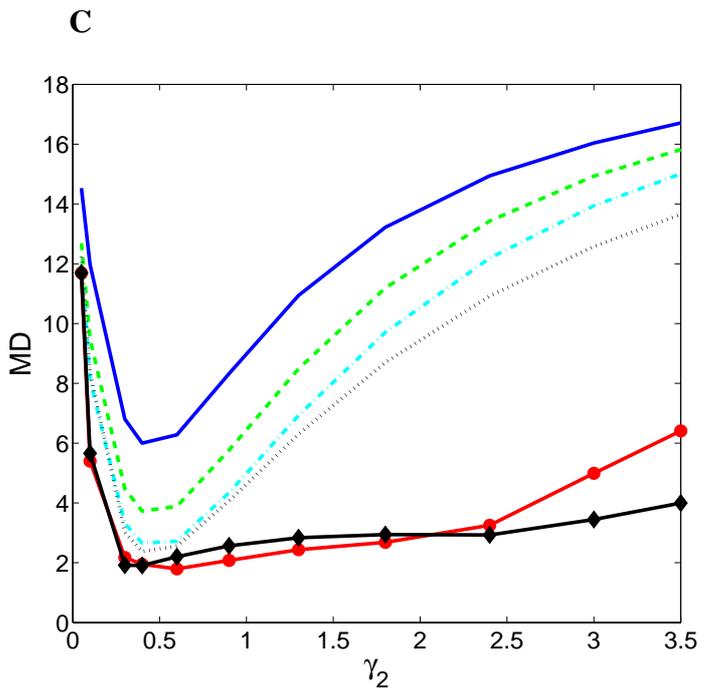
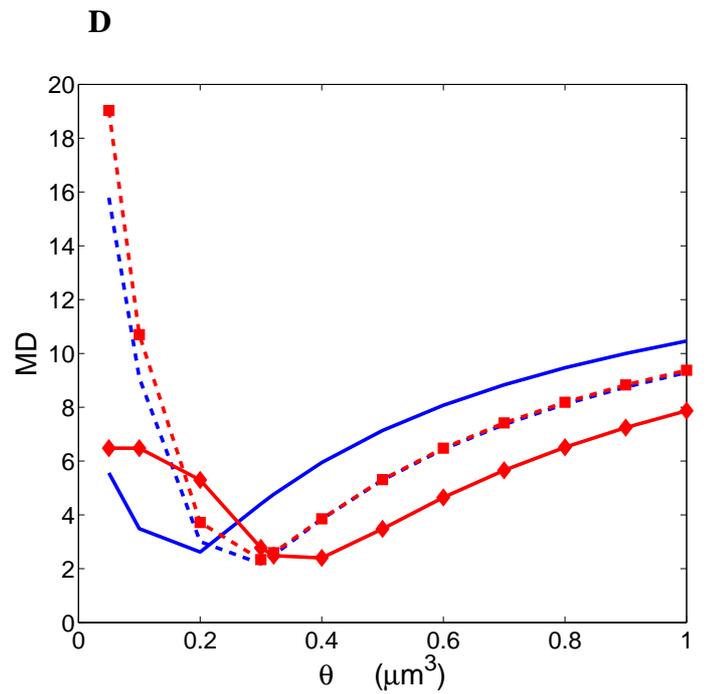
{\bf Fig. 6}  \\
{\bf The same as in Fig. 5 but for the more general Mahalanobis distance (MD) between 
optimal theoretical fractions and empirical fractions.} 
(A) MD as a function of $r$, and (B) MD as a function of $\theta$ for wire minimization.
(C) MD as a function of $\gamma_{2}$, and (D) MD as a function of $\theta$ for spine 
economical maximization. In panel (D) $\gamma_{2}=0.2$ 
for solid lines and $\gamma_{2}=0.5$ for dashed lines.

\vspace{0.3cm}

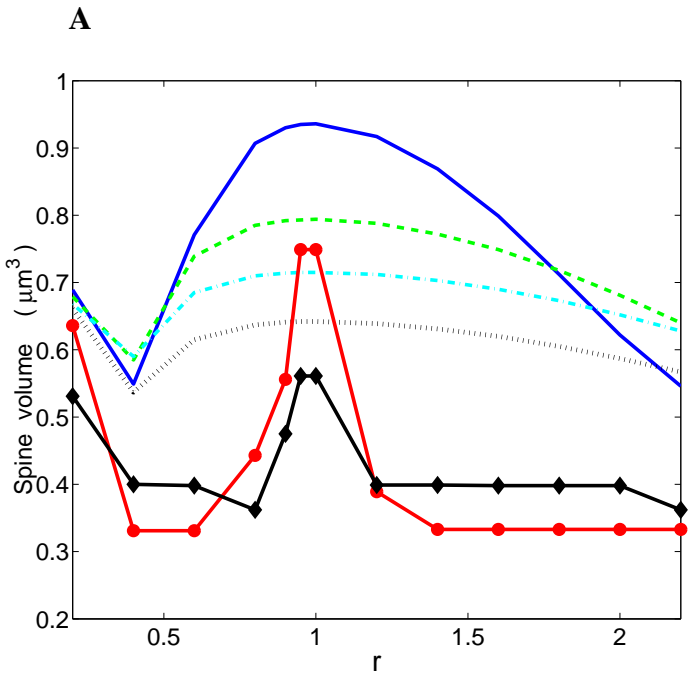
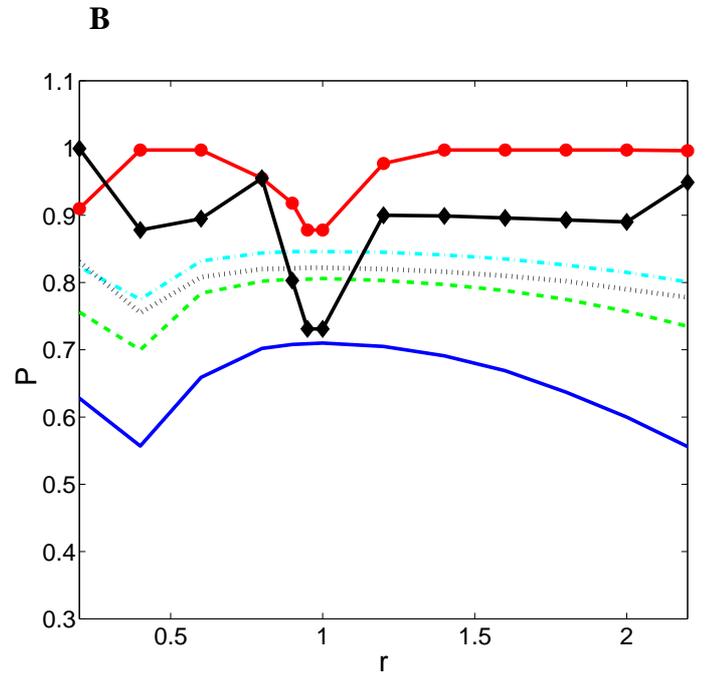
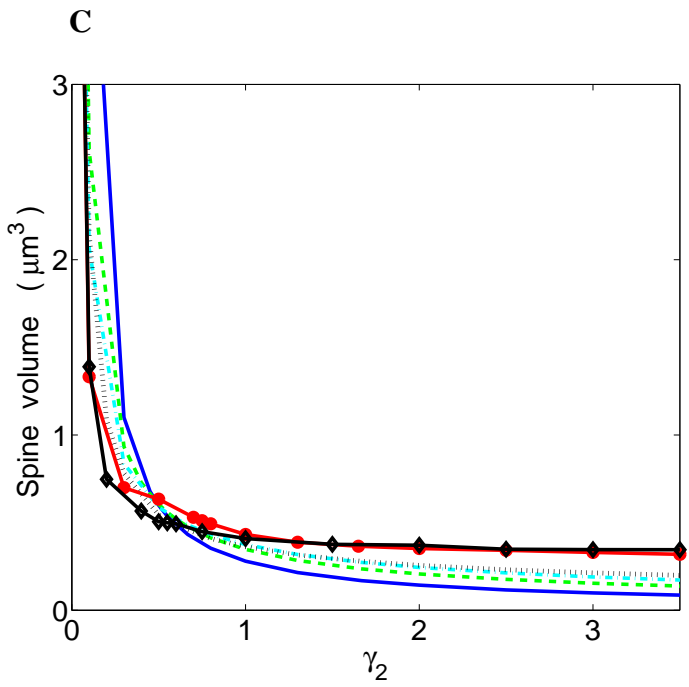
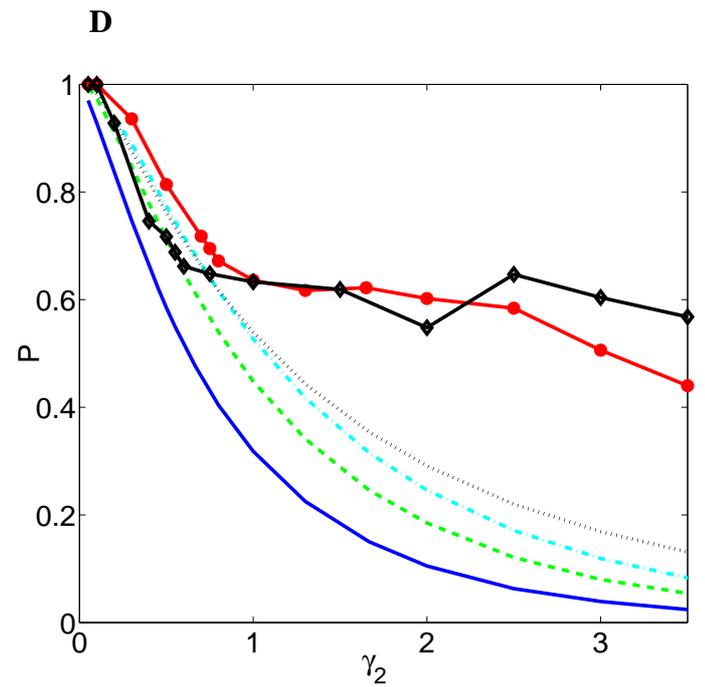
{\bf Fig. 7} \\
{\bf Optimal average spine volume $\overline{u}$ and conditional probability
of spine formation $P$ for different distributions of spine sizes.}
Panels (A) and (B) refer to ``wire volume minimization'', whereas panels (C) and
(D) correspond to ``spine economical maximization''.
(A) Non-monotonic dependence of spine volume $\overline{u}$ and (B)
conditional probability $P$ on $r$ for wire fractional volume minimization
($\gamma_{1}= 0$).
(C) Spine volume  $\overline{u}$ and (D) conditional probability $P$ 
decrease monotonically with increasing the exponent $\gamma_{2}$.
For all panels the same labels for curves corresponding to a given 
distribution were used, and they are identical to the labels used in 
Figs. 5 A, C. 
The curves for Log-logistic and Log-normal correspond to different values
of the parameters, respectively, $\beta$ and $\sigma$ that yield the minimal
ED for a given $r$ or $\gamma_{2}$.
For all panels $\theta= 0.321$ $\mu$m$^{3}$.

\vspace{0.3cm}

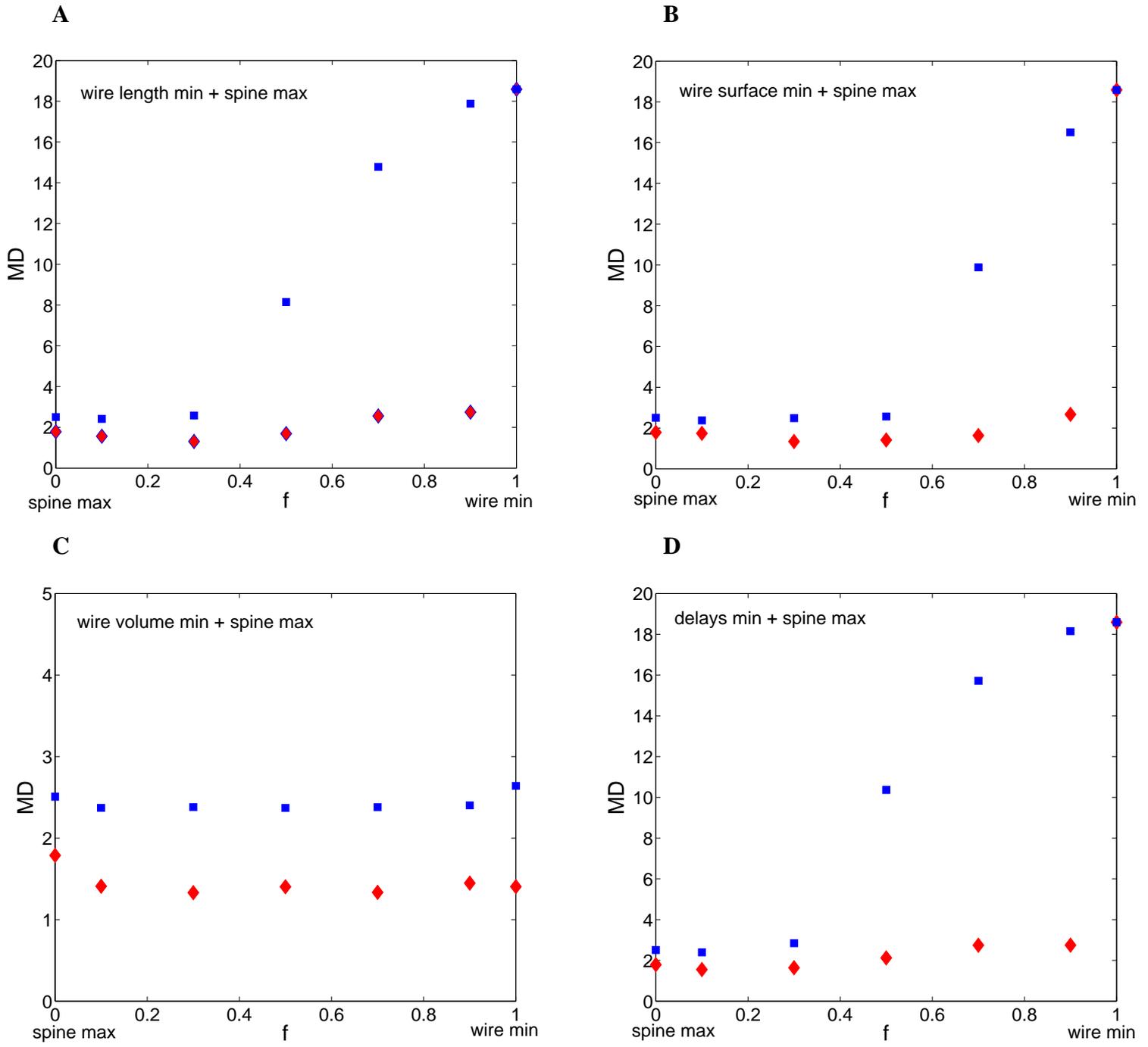
{\bf Fig. 8}  \\
{\bf The minimal value of MD as a function the control parameter $f$ for the combined
``spine economy max and wire min'' principle.}
Blue squares correspond to Gamma (n=2) distribution of spine sizes and red diamonds to 
Log-normal distribution. The notion of spine economy max is mixed with different types of wire 
cost min principle: wire length in (A), wire surface area in (B), wire volume in (C), and 
conduction delays in (D). In almost all panels MD has extremely shallow minima for $f \sim 0.1-0.3$, 
and then MD increases either weakly or abruptly as $f$ increases further (for $f=1$ MD
is the same for both distributions of spine sizes). The exception is the mixture of wire 
volume min and spine max (panel C), where MD is practically constant. 
For all panels $\theta= 0.321$ $\mu$m$^{3}$.

\newpage

\begin{table}
\begin{center}
\caption{ Structural composition of the gray matter of cerebral cortex.} 

\begin{tabular}{|l l l l l l|}
\hline

Species      &\multicolumn{5}{c |}{Volume fraction ($\%$)} \\   
             &   Axons     &   Dendrites   &   Spines      &    Glia/            &  Capillaries     \\
             &             &               &               &    Astrocytes       &              \\

\hline

Mouse        &  34.0            &     35.0         &     14.0        &       11.0$^{*}$    &  $0.7 \pm 0.1$   \\
Rat          &  $47.0 \pm 5.0$  &  $35.0 \pm 5.0$  &      9.0        &  $8.0 \pm 4.0^{*}$  &     1.4          \\
Rabbit       &  $47.0 \pm 5.5 $ &  $34.7 \pm 3.9$  &  $5.7 \pm 0.8$  &  $12.7 \pm 2.2^{*}$ &     $-$          \\
Cat          &  $27.8 \pm 5.7$  &  $31.0 \pm 6.3$  &    $-$          &    15.5             &  $2.1 \pm 0.5$   \\    
Macaque      &  $-$             &  $33.0 \pm 19.0$ &  $4.5 \pm 0.8$  &    $-$              &  $0.9 \pm 0.1$   \\
Human        &  $-$             &  $35.4 \pm 23.7$ &  $14.8 \pm 9.9$ &  $11.5 \pm 3.4$     &  $1.7 \pm 0.3$   \\
             &                  &                  &                 &                     &                  \\
Species mean  & $39.0 \pm 2.3$  &  $34.0 \pm 5.3$  &  $9.6 \pm 2.0$  &  $11.7 \pm 1.1$     &  $1.4 \pm 0.1$   \\
Normalized mean & $40.8 \pm 2.4$ & $35.5 \pm 5.5$  &  $10.0 \pm 2.1$ &  $12.2 \pm 1.2$     &  $1.5 \pm 0.1$   \\
Rule ``powers of 1/3'' & 33.3    &  33.3           &    11.1         &     11.1            &    1.2           \\  

\hline

\hline
\end{tabular}
\end{center}
\end{table} 

\noindent
Symbol $^{*}$ corresponds to the fraction of unspecified glia cells. 
The next-to-last line contains normalized to 100$\%$ mean fractional values over species.
The last line is a theoretical prediction based on a ``powers of 1/3'' rule.
References for the data are given in the Methods.

\newpage

\begin{table}
\begin{center}
\caption{ The best optimal theoretical fractional volumes and related parameters in the 
cortex for a particular case of ``wire minimization'' principle associated with wire volume 
minimization ($\gamma_{1}=0$). The optimal fractions correspond either to minimal Euclidean 
distance (ED) or minimal Mahalanobis distance (MD) between theory and data (given in bold face).}

\begin{tabular}{|l c c c c c c c c c c c|}
\hline

Spine size   & $\theta$  &\multicolumn{8}{c}{Optimal parameters}  &  ED  &    MD    \\
distribution &           &  $x$  &  $y$  &  $s$   & $g$   &  $c$  & $\overline{u}$ & $P$ & $r$  &    &    \\

\hline

Exponential  &  0.100   &  0.388 & 0.330 & 0.068 & 0.201 & 0.014 &   0.157  &  0.528 &  0.94 & {\bf 0.091} & 6.981  \\
             &  0.100   &  0.403 & 0.316 & 0.067 & 0.200 & 0.013 &   0.157  &  0.528 &  0.91 &  0.094      & {\bf 6.942}  \\
             &  0.321   &  0.423 & 0.371 & 0.111 & 0.085 & 0.009 &   0.935  &  0.709 &  0.96 & {\bf 0.045} & 6.485  \\
             &  0.321   &  0.397 & 0.397 & 0.112 & 0.085 & 0.010 &   0.936  &  0.710 &  1.00 &  0.058      & {\bf 6.360} \\
             &          &        &       &       &       &       &          &        &       &             &         \\
Gamma        &  0.100   &  0.376 & 0.316 & 0.081 & 0.210 & 0.017 &   0.175  &  0.684 &  0.93 & {\bf 0.104} & 7.842  \\
$\;$ (n=1)   &  0.100   &  0.442 & 0.259 & 0.078 & 0.206 & 0.016 &   0.175  &  0.682 &  0.80 & 0.134       & {\bf 7.427}  \\
             &  0.321   &  0.413 & 0.357 & 0.119 & 0.099 & 0.012 &   0.793  &  0.805 &  0.95 & {\bf 0.030} & 3.991  \\
             &  0.321   &  0.396 & 0.374 & 0.119 & 0.099 & 0.012 &   0.794  &  0.806 &  0.98 & 0.037       & {\bf 3.892} \\
             &          &        &       &       &       &       &          &        &       &             &          \\
Gamma        &  0.100   &  0.366 & 0.310 & 0.086 & 0.219 & 0.019 &   0.175  &  0.753 &  0.93 & {\bf 0.115} & 9.078  \\
$\;$ (n=2)   &  0.100   &  0.463 & 0.230 & 0.080 & 0.210 & 0.017 &   0.174  &  0.750 &  0.74 & 0.163       & {\bf 8.173}  \\
             &  0.321   &  0.406 & 0.352 & 0.121 & 0.108 & 0.013 &   0.715  &  0.846 &  0.95 & {\bf 0.026} & 2.681   \\
             &  0.321   &  0.395 & 0.363 & 0.121 & 0.108 & 0.013 &   0.715  &  0.846 &  0.97 & 0.030       & {\bf 2.640}  \\
             &          &        &       &       &       &       &          &        &       &             &          \\
Rayleigh     &  0.100   &  0.368 & 0.305 & 0.082 & 0.226 & 0.019 &   0.159  &  0.734 &  0.92 & {\bf 0.124} & 9.623   \\
             &  0.100   &  0.471 & 0.221 & 0.076 & 0.216 & 0.016 &   0.158  &  0.731 &  0.72 &  0.177      & {\bf 8.701} \\
             &  0.321   &  0.405 & 0.352 & 0.117 & 0.113 & 0.013 &   0.642  &  0.822 &  0.95 & {\bf 0.020} & 2.271    \\
             &  0.321   &  0.394 & 0.362 & 0.117 & 0.113 & 0.013 &   0.642  &  0.822 &  0.97 & 0.025       & {\bf 2.236} \\
             &          &        &       &       &       &       &          &        &       &             &         \\
Log-logistic &  0.100   &  0.404 & 0.350 & 0.097 & 0.136 & 0.013 &   0.404  &  0.683 &  0.95 & {\bf 0.015} & 2.378  \\
             &  0.100   &  0.399 & 0.356 & 0.097 & 0.136 & 0.013 &   0.404  &  0.684 &  0.96 &  0.017      & {\bf 2.351}  \\
             &  0.321   &  0.399 & 0.348 & 0.124 & 0.114 & 0.014 &   0.671  &  0.891 &  0.95 & {\bf 0.027} & 1.803  \\
             &  0.321   &  0.403 & 0.334 & 0.123 & 0.124 & 0.015 &   0.584  &  0.910 &  0.93 &  0.031      & {\bf 1.448} \\
             &          &        &       &       &       &       &          &        &       &             &          \\
Log-normal   &  0.100   &  0.418 & 0.367 & 0.072 & 0.133 & 0.010 &   0.311  &  0.472 &  0.96 & {\bf 0.034} &  5.734   \\
             &  0.100   &  0.401 & 0.356 & 0.073 & 0.158 & 0.012 &   0.244  &  0.513 &  0.96 &  0.045      & {\bf 4.801} \\
             &  0.321   &  0.411 & 0.354 & 0.106 & 0.116 & 0.012 &   0.561  &  0.731 &  0.95 & {\bf 0.010} & 2.897  \\
             &  0.321   &  0.402 & 0.341 & 0.110 & 0.133 & 0.015 &   0.475  &  0.804 &  0.94 &  0.021      & {\bf 1.405} \\

\hline

\hline
\end{tabular}
\end{center}

For log-logistic distribution the minimal ED and MD were obtained for the parameter $\beta=1.5$ if 
$\theta=0.100$, and if $\theta=0.321$ then $\beta= 3.5$ for minimal ED and $\beta=4.5$ for minimal MD. 
For log-normal distribution the minimal ED was reached for $\sigma=0.7$ and minimal MD for $\sigma=0.55$ if $\theta=0.100$,
and if $\theta=0.321$ then $\sigma=0.3$ for minimal ED and $\sigma=0.2$ for minimal MD. 

\end{table}

\newpage

\begin{table}
\begin{center}
\caption{ The best optimal theoretical fractional volumes and related parameters in the 
cortex for ``spine economical maximization'' principle. The optimal fractions correspond to
either the minimal Euclidean distance (ED) or Mahalanobis distance (MD) between theory and data
(given in bold face).}

\begin{tabular}{|l c c c c c c c c c c c|}
\hline

Spine size   & $\theta$  &\multicolumn{8}{c}{Optimal parameters}  &  ED  &  MD   \\
distribution &           &  $x$  &  $y$  &  $s$   & $g$   &  $c$  & $\overline{u}$  & $P$  & $\gamma_{2}$ &    &    \\

\hline

Exponential  &  0.100   &  0.374 & 0.374 & 0.119 & 0.118 & 0.014 &  0.615  &  0.850 &  0.25  & {\bf 0.043}  & 1.982   \\
             &  0.100   &  0.374 & 0.374 & 0.119 & 0.118 & 0.014 &  0.615  &  0.850 &  0.25  &  0.043  & {\bf 1.982}   \\
             &  0.321   &  0.398 & 0.398 & 0.093 & 0.102 & 0.009 &  0.599  &  0.585 &  0.50  & {\bf 0.050}  &  5.913  \\
             &  0.321   &  0.397 & 0.397 & 0.098 & 0.097 & 0.010 &  0.678  &  0.623 &  0.45  &  0.051  & {\bf 5.883}  \\
             &          &        &       &       &       &       &         &        &        &         &             \\
Gamma        &  0.100   &  0.366 & 0.366 & 0.129 & 0.123 & 0.016 &  0.626  &  0.959 &  0.20  & {\bf 0.051} & 2.353  \\
$\;$ (n=1)   &  0.100   &  0.366 & 0.366 & 0.129 & 0.123 & 0.016 &  0.626  &  0.959 &  0.20  &  0.051  & {\bf 2.353} \\
             &  0.321   &  0.388 & 0.388 & 0.098 & 0.116 & 0.011 &  0.520  &  0.650 &  0.60  & {\bf 0.039} & 3.886  \\
             &  0.321   &  0.385 & 0.385 & 0.111 & 0.107 & 0.012 &  0.660  &  0.746 &  0.45  &  0.042   & {\bf 3.597}  \\
             &          &        &       &       &       &       &         &        &        &          &            \\
Gamma        &  0.100   &  0.370 & 0.370 & 0.136 & 0.110 & 0.015 &  0.778  &  0.993 &  0.15  & {\bf 0.056}  & 2.554  \\
$\;$ (n=2)   &  0.100   &  0.370 & 0.370 & 0.136 & 0.110 & 0.015 &  0.778  &  0.993 &  0.15  &  0.056  & {\bf 2.554}  \\
             &  0.321   &  0.382 & 0.382 & 0.101 & 0.122 & 0.012 &  0.495  &  0.692 &  0.65  & {\bf 0.038}  & 2.914    \\
             &  0.321   &  0.380 & 0.380 & 0.112 & 0.116 & 0.013 &  0.589  &  0.774 &  0.50  &  0.040  & {\bf 2.507}   \\
             &          &        &       &       &       &       &         &        &        &         &              \\
Rayleigh     &  0.100   &  0.361 & 0.361 & 0.127 & 0.135 & 0.017 &  0.534  &  0.973 &  0.20  & {\bf 0.056} & 3.306     \\
             &  0.100   &  0.371 & 0.371 & 0.136 & 0.108 & 0.015 &  0.806  &  0.988 &  0.15  &   0.056 & {\bf 2.645}  \\
             &  0.321   &  0.380 & 0.380 & 0.102 & 0.125 & 0.013 &  0.486  &  0.710 &  0.60  & {\bf 0.038} &  2.555   \\
             &  0.321   &  0.378 & 0.378 & 0.113 & 0.118 & 0.013 &  0.585  &  0.789 &  0.45  &  0.040  & {\bf 2.284}   \\
             &          &        &       &       &       &       &         &        &        &         &             \\
Log-logistic &  0.100   &  0.377 & 0.377 & 0.106 & 0.126 & 0.013 &  0.498  &  0.747 &  0.40  & {\bf 0.039}  &  2.152  \\
             &  0.100   &  0.377 & 0.377 & 0.106 & 0.126 & 0.013 &  0.498  &  0.747 &  0.40  &  0.039  & {\bf 2.152}   \\
             &  0.321   &  0.383 & 0.383 & 0.102 & 0.120 & 0.012 &  0.511  &  0.695 &  0.75  & {\bf 0.038}  & 2.999   \\
             &  0.321   &  0.372 & 0.372 & 0.114 & 0.127 & 0.015 &  0.524  &  0.824 &  0.60  &  0.043  & {\bf 1.793}  \\
             &          &        &       &       &       &       &         &        &        &         &             \\
Log-normal   &  0.100   &  0.381 & 0.381 & 0.098 & 0.128 & 0.013 &  0.445  &  0.675 &  0.30  & {\bf 0.038}  &  2.794  \\
             &  0.100   &  0.372 & 0.372 & 0.120 & 0.120 & 0.015 &  0.603  &  0.868 &  0.20  &  0.045  & {\bf 1.880}   \\
             &  0.321   &  0.383 & 0.383 & 0.101 & 0.121 & 0.012 &  0.499  &  0.688 &  0.55  & {\bf 0.038} & 3.005   \\
             &  0.321   &  0.372 & 0.372 & 0.115 & 0.126 & 0.014 &  0.535  &  0.828 &  0.35  & 0.043  & {\bf 1.788}    \\

\hline

\hline
\end{tabular}
\end{center}

For log-logistic distribution the minimal ED and MD were obtained for the parameter $\beta=1.5$ if 
$\theta=0.100$, and if $\theta=0.321$ then $\beta=3.0$ for minimal ED and $\beta=4.0$ for minimal MD. 
For log-normal distribution the minimal ED and MD were reached for $\sigma=0.75$ if $\theta=0.100$, and
if $\theta=0.321$ then $\sigma=0.25$ for minimal ED and MD. 

\end{table}

\newpage

\begin{table}
\begin{center}
\caption{ The best optimal theoretical fractional volumes and related parameters in the 
cortex for the combined ``wire min + spine max'' principle with control parameter $f=0.1$. 
The optimal fractions correspond to the minimal Mahalanobis distance (MD) between theory and data.} 

\begin{tabular}{|l c c c c c c c c c c |}
\hline

Principle type/        &\multicolumn{9}{c}{Optimal parameters}                                          &   MD  \\
 $\; \;$ spine distr.  &  $x$  &  $y$  &  $s$  & $g$ & $c$ & $\overline{u}$ & $P$  & $r$  & $\gamma_{2}$ &       \\

\hline
                     &        &       &       &       &       &         &         &        &       &        \\
\multicolumn{5}{|l}{{\bf Wire length min + spine max} ($\gamma_{1}=2/3$)}     & & & & & &      \\ 
$\; \;$ Exponential  &  0.399 & 0.396 & 0.097 & 0.098 & 0.010 &  0.655  &  0.612  &  0.99  & 1.00  &  5.885    \\
$\; \;$ Gamma (n=1)  &  0.395 & 0.377 & 0.107 & 0.110 & 0.012 &  0.612  &  0.717  &  0.90  & 0.95  &  3.558   \\
$\; \;$ Gamma (n=2)  &  0.393 & 0.368 & 0.110 & 0.117 & 0.013 &  0.569  &  0.759  &  0.85  & 0.95  &  2.416   \\
$\; \;$ Rayleigh     &  0.395 & 0.362 & 0.110 & 0.120 & 0.013 &  0.553  &  0.768  &  0.80  & 0.90  &  2.117     \\
$\; \;$ Log-logistic &  0.400 & 0.344 & 0.116 & 0.126 & 0.015 &  0.544  &  0.844  &  0.65  & 0.90  &  1.277    \\
$\; \;$ Log-normal   &  0.386 & 0.353 & 0.112 & 0.134 & 0.015 &  0.480  &  0.825  &  0.80  & 0.80  &  1.565   \\
                     &        &       &       &       &       &         &         &        &       &           \\
\multicolumn{5}{|l}{{\bf Wire surface min + spine max} ($\gamma_{1}=1/3$)}      & & & & & &      \\ 
$\; \;$ Exponential  &  0.402 & 0.393 & 0.100 & 0.096 & 0.010 &  0.696  &  0.631  &  0.95  & 0.70  &  5.903    \\
$\; \;$ Gamma (n=1)  &  0.394 & 0.377 & 0.108 & 0.109 & 0.012 &  0.629  &  0.728  &  0.90  & 0.70  &  3.549     \\
$\; \;$ Gamma (n=2)  &  0.396 & 0.364 & 0.111 & 0.116 & 0.013 &  0.589  &  0.774  &  0.80  & 0.70  &  2.375    \\
$\; \;$ Rayleigh     &  0.399 & 0.358 & 0.111 & 0.118 & 0.013 &  0.576  &  0.784  &  0.75  & 0.65  &  2.105     \\
$\; \;$ Log-logistic &  0.398 & 0.346 & 0.117 & 0.125 & 0.015 &  0.550  &  0.850  &  0.65  & 0.70  &  1.280    \\
$\; \;$ Log-normal   &  0.396 & 0.347 & 0.104 & 0.138 & 0.014 &  0.423  &  0.759  &  0.65  & 0.80  &  1.741    \\
                     &        &       &       &       &       &         &         &        &       &            \\
\multicolumn{5}{|l}{{\bf Wire volume min + spine max} ($\gamma_{1}=0 $)}      & & & & & &       \\ 
$\; \;$ Exponential  &  0.399 & 0.396 & 0.099 & 0.096 & 0.010 &  0.692  &  0.629  &  0.98  & 0.45  &  5.898    \\
$\; \;$ Gamma (n=1)  &  0.398 & 0.374 & 0.107 & 0.110 & 0.012 &  0.617  &  0.721  &  0.85  & 0.50  &  3.553     \\
$\; \;$ Gamma (n=2)  &  0.395 & 0.365 & 0.112 & 0.115 & 0.013 &  0.598  &  0.781  &  0.80  & 0.50  &  2.371    \\
$\; \;$ Rayleigh     &  0.397 & 0.359 & 0.113 & 0.117 & 0.013 &  0.590  &  0.792  &  0.75  & 0.45  &  2.111     \\
$\; \;$ Log-logistic &  0.399 & 0.345 & 0.117 & 0.125 & 0.015 &  0.548  &  0.848  &  0.60  & 0.55  &  1.277    \\
$\; \;$ Log-normal   &  0.400 & 0.343 & 0.110 & 0.133 & 0.015 &  0.474  &  0.798  &  0.60  & 0.45  &  1.409    \\
                     &        &       &       &       &       &         &         &        &       &           \\
\multicolumn{5}{|l}{{\bf Delays min + spine max} ($\gamma_{1}=5/6 $)}        & & & & & &      \\ 
$\; \;$ Exponential  &  0.398 & 0.398 & 0.096 & 0.099 & 0.010 &  0.642  &  0.606  &  1.00  & 1.15  &  5.887    \\
$\; \;$ Gamma (n=1)  &  0.390 & 0.382 & 0.106 & 0.111 & 0.012 &  0.597  &  0.708  &  0.95  & 1.10  &  3.588     \\
$\; \;$ Gamma (n=2)  &  0.393 & 0.367 & 0.111 & 0.117 & 0.013 &  0.580  &  0.768  &  0.85  & 1.05  &  2.393    \\
$\; \;$ Rayleigh     &  0.392 & 0.366 & 0.108 & 0.121 & 0.013 &  0.537  &  0.755  &  0.85  & 1.05  &  2.164     \\
$\; \;$ Log-logistic &  0.396 & 0.348 & 0.114 & 0.127 & 0.014 &  0.530  &  0.830  &  0.70  & 1.05  &  1.295    \\
$\; \;$ Log-normal   &  0.390 & 0.356 & 0.108 & 0.132 & 0.014 &  0.469  &  0.775  &  0.80  & 1.00  &  1.553    \\
                     &        &       &       &       &       &         &         &        &       &          \\

\hline

\hline
\end{tabular}
\end{center}
All the results correspond to $\theta=0.321$.

\end{table}

\newpage

\begin{table}
\begin{center}
\caption{ The best optimal theoretical fractional volumes and related parameters in the 
cortex for the combined ``wire min + spine max'' principle with control parameter $f=0.5$. 
The optimal fractions correspond to the minimal Mahalanobis distance (MD) between theory and data.} 

\begin{tabular}{|l c c c c c c c c c c |}
\hline

Principle type/        &\multicolumn{9}{c}{Optimal parameters}                                          &   MD   \\
 $\; \;$ spine distr.  &  $x$  &  $y$  &  $s$  & $g$ & $c$ & $\overline{u}$ & $P$  & $r$  & $\gamma_{2}$ &        \\

\hline
                     &        &       &       &       &       &         &         &        &       &          \\
\multicolumn{5}{|l}{{\bf Wire length min + spine max} ($\gamma_{1}=2/3$)}   &  & & & & &      \\ 
$\; \;$ Exponential  &  0.453 & 0.453 & 0.013 & 0.080 & 0.001 &  0.115  &  0.061  &  1.00  & 2.95  &  15.22    \\
$\; \;$ Gamma (n=1)  &  0.421 & 0.433 & 0.036 & 0.107 & 0.004 &  0.212  &  0.195  &  1.05  & 2.75  &  11.75   \\
$\; \;$ Gamma (n=2)  &  0.399 & 0.413 & 0.059 & 0.121 & 0.007 &  0.290  &  0.356  &  1.05  & 2.60  &  8.157   \\
$\; \;$ Rayleigh     &  0.393 & 0.407 & 0.065 & 0.127 & 0.008 &  0.299  &  0.405  &  1.05  & 2.55  &  7.047     \\
$\; \;$ Log-logistic &  0.392 & 0.359 & 0.099 & 0.137 & 0.014 &  0.409  &  0.706  &  0.90  & 2.40  &  2.118    \\
$\; \;$ Log-normal   &  0.399 & 0.353 & 0.110 & 0.125 & 0.014 &  0.520  &  0.777  &  0.90  & 2.10  &  1.692   \\
                     &        &       &       &       &       &         &         &        &       &          \\
\multicolumn{5}{|l}{{\bf Wire surface min + spine max} ($\gamma_{1}=1/3$)}      &  & & & & &       \\ 
$\; \;$ Exponential  &  0.421 & 0.421 & 0.040 & 0.114 & 0.005 &  0.215  &  0.225  &  1.00  & 2.00  &  10.94    \\ 
$\; \;$ Gamma (n=1)  &  0.395 & 0.395 & 0.079 & 0.122 & 0.010 &  0.389  &  0.509  &  1.00  & 1.85  &  5.508   \\ 
$\; \;$ Gamma (n=2)  &  0.380 & 0.380 & 0.109 & 0.118 & 0.013 &  0.558  &  0.751  &  1.00  & 1.70  &  2.566   \\  
$\; \;$ Rayleigh     &  0.392 & 0.368 & 0.102 & 0.125 & 0.013 &  0.486  &  0.710  &  0.95  & 1.65  &  2.436   \\   
$\; \;$ Log-logistic &  0.404 & 0.336 & 0.114 & 0.131 & 0.015 &  0.500  &  0.835  &  0.85  & 1.60  &  1.350    \\
$\; \;$ Log-normal   &  0.395 & 0.347 & 0.110 & 0.133 & 0.015 &  0.475  &  0.804  &  0.90  & 1.40  &  1.414   \\
                     &        &       &       &       &       &         &         &        &       &          \\
\multicolumn{5}{|l}{{\bf Wire volume min + spine max} ($\gamma_{1}=0 $)}      &  & & & & &      \\ 
$\; \;$ Exponential  &  0.398 & 0.398 & 0.097 & 0.098 & 0.010 &  0.655  &  0.613  &  1.00  & 0.55  &  5.886    \\ 
$\; \;$ Gamma (n=1)  &  0.403 & 0.369 & 0.107 & 0.109 & 0.012 &  0.617  &  0.721  &  0.95  & 0.60  &  3.566    \\  
$\; \;$ Gamma (n=2)  &  0.395 & 0.364 & 0.112 & 0.115 & 0.013 &  0.598  &  0.781  &  0.95  & 0.60  &  2.370    \\  
$\; \;$ Rayleigh     &  0.394 & 0.363 & 0.111 & 0.119 & 0.013 &  0.567  &  0.777  &  0.95  & 0.55  &  2.107    \\   
$\; \;$ Log-logistic &  0.400 & 0.344 & 0.116 & 0.126 & 0.015 &  0.544  &  0.844  &  0.90  & 0.70  &  1.278    \\ 
$\; \;$ Log-normal   &  0.401 & 0.340 & 0.111 & 0.133 & 0.015 &  0.478  &  0.816  &  0.90  & 0.40  &  1.403   \\
                     &        &       &       &       &       &         &         &        &       &          \\
\multicolumn{5}{|l}{{\bf Delays min + spine max} ($\gamma_{1}=5/6 $)}      &  & & & & &      \\ 
$\; \;$ Exponential  &  0.463 & 0.463 & 0.008 & 0.065 & 0.001 &  0.097  &  0.036  &  1.00  & 3.40  &  16.15   \\
$\; \;$ Gamma (n=1)  &  0.438 & 0.438 & 0.026 & 0.096 & 0.002 &  0.183  &  0.135  &  1.00  & 3.10  &  13.34   \\ 
$\; \;$ Gamma (n=2)  &  0.419 & 0.419 & 0.045 & 0.113 & 0.005 &  0.249  &  0.257  &  1.00  & 2.95  &  10.36   \\ 
$\; \;$ Rayleigh     &  0.410 & 0.410 & 0.053 & 0.121 & 0.006 &  0.265  &  0.317  &  1.00  & 2.90  &  8.918   \\ 
$\; \;$ Log-logistic &  0.396 & 0.345 & 0.105 & 0.139 & 0.015 &  0.420  &  0.765  &  0.85  & 2.50  &  1.761    \\    
$\; \;$ Log-normal   &  0.405 & 0.337 & 0.100 & 0.144 & 0.014 &  0.383  &  0.732  &  0.80  & 2.40  &  2.124   \\
                     &        &       &       &       &       &         &         &        &       &           \\

\hline

\hline
\end{tabular}
\end{center}
All the results correspond to $\theta=0.321$.

\end{table}

\newpage

\begin{table}
\begin{center}
\caption{ The best optimal theoretical fractional volumes and related parameters in the 
cortex for the combined ``wire min + spine max'' principle with control parameter $f=0.9$. 
The optimal fractions correspond to the minimal Mahalanobis distance (MD) between theory and data.} 

\begin{tabular}{|l c c c c c c c c c c |}
\hline

Principle type/        &\multicolumn{9}{c}{Optimal parameters}                                          &   MD   \\
 $\; \;$ spine distr.  &  $x$  &  $y$  &  $s$  & $g$ & $c$ & $\overline{u}$ & $P$  & $r$  & $\gamma_{2}$ &        \\

\hline 
                     &        &        &       &       &       &       &         &        &        &           \\
\multicolumn{5}{|l}{{\bf Wire length min + spine max} ($\gamma_{1}=2/3$)}  &  & & & & &      \\ 
$\; \;$ Exponential  &  0.414 & 0.414 & 0.172 & 0.000 & 0.000 &  22077.8 & 1.000  &  1.00  & 0.10  &  18.45   \\
$\; \;$ Gamma (n=1)  &  0.489 & 0.489 & 0.001 & 0.020 & 0.000 &  0.086  &  0.005  &  1.00  & 6.40  &  18.33  \\
$\; \;$ Gamma (n=2)  &  0.484 & 0.484 & 0.003 & 0.029 & 0.000 &  0.118  &  0.012  &  1.00  & 6.20  &  17.88  \\
$\; \;$ Rayleigh     &  0.471 & 0.471 & 0.008 & 0.050 & 0.000 &  0.157  &  0.038  &  1.00  & 6.10  &  16.69   \\
$\; \;$ Log-logistic &  0.414 & 0.414 & 0.172 & 0.000 & 0.000 &  22073.2 & 1.000  &  1.00  & 0.10  &  18.45   \\
$\; \;$ Log-normal   &  0.401 & 0.340 & 0.093 & 0.151 & 0.014 &  0.332  &  0.684  &  0.35  & 5.50  &  2.751   \\
                     &        &        &       &       &       &       &         &        &        &           \\
\multicolumn{5}{|l}{{\bf Wire surface min + spine max} ($\gamma_{1}=1/3$)} &  &  & & & &                       \\ 
$\; \;$ Exponential  &  0.487 & 0.487 & 0.001 & 0.025 & 0.000 &  0.060  &  0.005  &  1.00  & 5.20  &  18.13   \\
$\; \;$ Gamma (n=1)  &  0.478 & 0.478 & 0.004 & 0.040 & 0.000 &  0.107  &  0.018  &  1.00  & 4.90  &  17.38   \\
$\; \;$ Gamma (n=2)  &  0.468 & 0.468 & 0.009 & 0.054 & 0.001 &  0.146  &  0.040  &  1.00  & 4.65  &  16.51  \\
$\; \;$ Rayleigh     &  0.454 & 0.454 & 0.017 & 0.074 & 0.001 &  0.181  &  0.084  &  1.00  & 4.60  &  15.07   \\
$\; \;$ Log-logistic &  0.402 & 0.339 & 0.095 & 0.150 & 0.014 &  0.341  &  0.696  &  0.75  & 4.20  &  2.622    \\
$\; \;$ Log-normal   &  0.402 & 0.347 & 0.092 & 0.145 & 0.013 &  0.348  &  0.664  &  0.85  & 3.80  &  2.670   \\
                     &        &       &       &       &       &         &         &        &      &        \\
\multicolumn{5}{|l}{{\bf Wire volume min + spine max} ($\gamma_{1}=0 $)} &  & & & & &       \\ 
$\; \;$ Exponential  &  0.398 & 0.398 & 0.097 & 0.099 & 0.010 &  0.651  &  0.611  &  1.00  & 1.10  &  5.886  \\
$\; \;$ Gamma (n=1)  &  0.386 & 0.386 & 0.108 & 0.109 & 0.012 &  0.628  &  0.728  &  1.00  & 1.20  &  3.591  \\ 
$\; \;$ Gamma (n=2)  &  0.404 & 0.355 & 0.112 & 0.115 & 0.013 &  0.601  &  0.783  &  0.95  & 1.20  &  2.401  \\
$\; \;$ Rayleigh     &  0.403 & 0.354 & 0.111 & 0.119 & 0.013 &  0.568  &  0.779  &  0.95  & 1.00  &  2.126  \\
$\; \;$ Log-logistic &  0.394 & 0.350 & 0.114 & 0.127 & 0.015 &  0.529  &  0.829  &  0.95  & 1.60  &  1.311   \\
$\; \;$ Log-normal   &  0.396 & 0.349 & 0.109 & 0.132 & 0.014 &  0.472  &  0.788  &  0.95  & 0.70  &  1.447   \\
                     &        &       &       &       &       &         &         &        &       &           \\
\multicolumn{5}{|l}{{\bf Delays min + spine max} ($\gamma_{1}=5/6 $)} &  & & & & &     \\ 
$\; \;$ Exponential  &  0.425 & 0.403 & 0.171 & 0.001 & 0.000 & 2095.2  &  1.000  &  0.50  & 0.10  &  18.37   \\
$\; \;$ Gamma (n=1)  &  0.425 & 0.403 & 0.171 & 0.001 & 0.000 & 2091.7  &  1.000  &  0.50  & 0.10  &  18.37   \\
$\; \;$ Gamma (n=2)  &  0.487 & 0.487 & 0.002 & 0.024 & 0.000 &  0.111  &  0.008  &  1.00  & 6.70  &  18.15  \\
$\; \;$ Rayleigh     &  0.475 & 0.475 & 0.006 & 0.043 & 0.000 &  0.151  &  0.029  &  1.00  & 6.70  &  17.11   \\
$\; \;$ Log-logistic &  0.425 & 0.403 & 0.171 & 0.001 & 0.000 & 2091.7  &  1.000  &  0.50  & 0.10  &  18.37   \\
$\; \;$ Log-normal   &  0.401 & 0.340 & 0.093 & 0.151 & 0.014 &  0.332  &  0.684  &  0.40  & 5.60  &  2.751   \\
                     &        &       &       &       &       &         &         &        &       &          \\

\hline

\hline
\end{tabular}
\end{center}
All the results correspond to $\theta=0.321$.

\end{table}


\begin{thebibliography}{99}

\bibitem{iadecola}
Iadecola C (2004) Neurovascular regulation in the normal brain and
in Alzheimer's disease. {\it Nat. Rev. Neurosci.} {\bf 5}: 347-360.

\bibitem{attwell2010}
Attwell D, Buchan AN, Charpak S, Lauritzen M, MacVicar BA, Newman EA
(2010) Glial and neuronal control of brain blood flow. {\it Nature}
{\bf 468}: 232-243.

\bibitem{wen2009}
Wen Q, Stepanyants A, Elston G, Grosberg AY, Chklovskii DB (2009)
Maximization of the connectivity repertoire as a statistical principle
governing the shapes of dendritic arbors. {\it Proc. Natl. Acad. Sci.
USA} {\bf 106}: 12536-12541.  

\bibitem{snider}
Snider J, Pillai A, Stevens CF (2010) A universal property of axonal
and dendritic arbors. {\it Neuron} {\bf 66}: 45-56.

\bibitem{cuntz}
Cuntz H, Mathy A, Hausser M (2012) A scaling law derived from optimal
dendritic wiring. {\it Proc. Natl. Acad. Sci. USA} {\bf 109}: 11014-11018.  


\bibitem{karbowski2011}
Karbowski J (2011) Scaling of brain metabolism and blood flow in relation
to capillary and neural scaling. {\it PLoS ONE} {\bf 6}: e26709.    

\bibitem{cherniak}
Cherniak C (1994) Component placement optimization in the
brain. {\it J. Neuroscience} {\bf 14}: 2418-2427.   

\bibitem{chklovskii}
Chklovskii DB, Schikorski T, Stevens CF (2002) Wiring optimization in
cortical circuits. {\it Neuron} {\bf 34}: 341-347.   

\bibitem{wen2005}
Wen Q, Chklovskii DB (2005) Segregation of the brain into gray and
white matter: A design minimizing conduction delays. {\it PLoS
Comput. Biol.} {\bf 1}: e78.  

\bibitem{klyachko}
Klyachko VA, Stevens CF (2003) Connectivity optimization and the positioning
of cortical areas. {\it Proc. Natl. Acad. Sci. USA} {\bf 100}: 7937-7941.

\bibitem{budd2010}
Budd JML, Kovacs K, Ferecsko AS, Buzas P, Eysel UT, Kisvarday ZF (2010)
Neocortical axon arbors trade-off material and conduction delay conservation.
{\it PLoS Comput. Biol.} {\bf 6}: e1000711.   

\bibitem{budd2012}
Budd JML, Kisvarday ZF (2012) Communication and wiring in the cortical connectome.
{\it Front. Neuroanat.} {\bf 6}: 42.   

\bibitem{bullmore}
Bullmore E, Sporns O (2012) The economy of brain network organization. {\it Nat.
Rev. Neurosci.}  {\bf 13}: 336-349.    


\bibitem{karbowski2014}
Karbowski J (2014) Constancy and trade-offs in the neuroanatomical and metabolic
design of the cerebral cortex. {\it Front. Neural Circuits} {\bf 8}: 9.

\bibitem{kaiser}
Kaiser M, Hilgetag CC (2006) Nonoptimal component placement, but short
processing paths, due to long-distance projections in neural systems.
{\it PLoS Comput. Biol.} {\bf 2}: e95.

\bibitem{chen}
Chen Y, Wang S, Hilgetag CC, Zhou C (2013) Trade-off between multiple constraints
enables simultaneous formation of modules and hubs in neural systems.
{\it PLoS Comput. Biol.} {\bf 9}: e1002937.

\bibitem{cherniak2004}
Cherniak C, Mokhtarzada Z, Rodriguez-Esteban R, Changizi K (2004) Global
optimization of cerebral cortex layout. {\it Proc. Natl. Acad. Sci. USA}
{\bf 101}: 1081-1086.


\bibitem{zecevic}
Zecevic N, Rakic P (1991) Synaptogenesis in monkey somatosensory cortex.
{\it Cereb. Cortex} {\bf 1}: 510-523.   


\bibitem{kasai}
Kasai H, Matsuzaki M, Noguchi J, Yasumatsu N, Nakahara H (2003) 
Structure-stability-function relationships of dendritic spines.
{\it Trends Neurosci.} {\bf 26}: 360-368.   

\bibitem{aiello}
Aiello LC, Wheeler P (1995) The expensive-tissue hypothesis: The brain
and the digestive-system in human and primate evolution. {\it Curr.
Anthropology} {\bf 36}: 199-221.   

\bibitem{attwell2001}
Attwell D, Laughlin SB (2001) An energy budget for signaling in the 
gray matter of the brain. {\it J. Cereb. Blood Flow Metabol.} 
{\bf 21}: 1133-1145.  

\bibitem{karbowski2007}
Karbowski J (2007) Global and regional brain metabolic scaling and its
functional consequences. {\it BMC Biology} {\bf 5}: 18.    


\bibitem{levy}
Levy WB, Baxter RA (1996) Energy efficient neural codes. {\it Neural
Computation} {\bf 8}: 531-543.   

\bibitem{laughlin}
Laughlin SB, de Ruyter van Steveninck RR, Anderson JC (1998) The metabolic
cost of neural information. {\it Nature Neurosci.} {\bf 1}: 36-40.   

\bibitem{alle}
Alle H, Roth A, Geiger JRP (2009) Energy-efficient action potentials
in hippocampal mossy fibers. {\it Science} {\bf 325}: 1405-1408.  

\bibitem{harris}
Harris JJ, Jolivet R, Attwell D (2012) Synaptic energy use and supply.
{\it Neuron} {\bf 75}: 762-777.   

\bibitem{karbowski2012}
Karbowski J (2012) Approximate invariance of metabolic energy per synapse
during development in mammalian brains. {\it PLoS ONE} {\bf 7}: e33425.   

\bibitem{chugani}
Chugani HT (1998) A critical period of brain development: Studies of
cerebral glucose utilization with PET. {\it Preventive Medicine} {\bf 27}:
184-188.   

\bibitem{tsai}
Tsai PS, Kaufhold J, Blinder P, Friedman B, Drew PJ et al (2009) 
Correlations of neuronal and microvascular densities
in murine cortex revealed by direct counting and colocalization of nuclei
and vessels. {\it J. Neurosci.} {\bf 29}: 14553-14570.   

\bibitem{weber}
Weber B, Keller AL, Reichold J, Logothetis NK (2008) The microvascular
system of the striate and extrastriate visual cortex of the macaque.
{\it Cereb. Cortex} {\bf 18}: 2318-2330. 

\bibitem{mccaslin}
McCaslin AFH, Chen BR, Radosevich AJ, Cauli B, Hillman EMC (2011)
In vivo 3D morphology of astrocyte-vasculature interactions in the
somatosensory cortex: implications for neurovascular coupling.
{\it J. Cereb. Blood Flow Metabol.} {\bf 31}: 795-806.  


\bibitem{montague}
Montague PR (1996) The resource consumption principle: attention and
memory in volumes of neural tissue. {\it Proc. Natl. Acad. Sci. USA}
{\bf 93}: 3619-3623.  

\bibitem{kaas}
Kaas JH (2000)  Why is brain size so important: Design
problems and solutions as neocortex gets bigger or smaller. {\it Brain Mind} 
{\bf 1}:  7-23.  

\bibitem{bassett}
Bassett DS, Greenfield DL, Meyer-Lindenberg A, Weinberger DR, Moore SW,
Bullmore ET (2010) Efficient physical embedding of topologically complex
information processing networks in brains and computer circuits.
{\it PLoS Comput. Biol.} {\bf 6}: e1000748.   


\bibitem{parker}
Parker GA, Maynard Smith J (1990) Optimality theory in evolutionary
biology. {\it Nature} {\bf 348}: 27-33.

\bibitem{alexander}
Alexander RM (1996) {\it Optima for Animals}. Princeton, NJ: Princeton
Univ. Press. 

\bibitem{striedter}
Striedter GF (2005) {\it Principles of Brain Evolution}. Sunderland, MA:
Sinauer Assoc.  


\bibitem{braitenberg}
Braitenberg V, Sch{\"u}z A (1998) {\it Cortex: Statistics
and Geometry of Neuronal Connectivity.} Berlin: Springer.   

\bibitem{houzel2010}
Herculano-Houzel S, Mota B, Wong P, Kaas JH (2010) 
Connectivity-driven white matter scaling and folding in primate cerebral
cortex. {\it Proc. Natl. Acad. Sci. USA} {\bf 107}: 19008-19013.

\bibitem{mayhew}
Mayhew TM, Mwamengele GLM, Dantzer V (1996) Stereological and allometric
studies on mammalian cerebral cortex with implications for medical brain
imaging. {\it J. Anat.} {\bf 189}: 177-184.

\bibitem{benavides}
Benavides-Piccione R, Fernaud-Espinosa I, Robles V, Yuste R, DeFelipe J
(2013) Age-based comparison of human dendritic spine structure using
complete three-dimensional reconstructions. {\it Cerebral Cortex}
{\bf 23}: 1798-1810.  

\bibitem{villalba}
Villalba RM, Smith Y (2010) Striatal spine plasticity in Parkinson's
disease. {\it Front. Neuroanat.} {\bf 4}: 133.


\bibitem{stevens}
Stevens CF (2011) Brain organization: Wiring economy works for the large
and small. {\it Curr. Biol.} {\bf 22}: R24. 


\bibitem{bonhoeffer}
Bonhoeffer T, Yuste R (2002) Spine motility: phenomenology, mechanisms,
and function. {\it Neuron} {\bf 35}: 1019-1027.   

\bibitem{holtmaat}
Holtmaat AJ, Trachtenberg JT, Wilbrecht L, Shepherd GM, Zhang X, et al 
(2005) Transient and persistent dendritic
spines in the neocortex in vivo. {\it Neuron} {\bf 45}: 279-291.   

\bibitem{statman}
Statman A, Kaufman M, Minerbi A, Ziv NE, Brenner N (2014) 
Synaptic size dynamics as an effective stochastic process.
{\it PLoS Comput. Biol.} {\bf 10}: e1003846.

\bibitem{meyer}
Meyer D, Bonhoeffer T, Scheuss V (2014) Balance and stability
of synaptic structures during synaptic plasticity. {\it Neuron}
{\bf 82}: 430-443.


\bibitem{hursh}
Hursh JB (1939) Conduction velocity and diameter of nerve fibers.
{\it Amer. J. Physiol.} {\bf 127}: 131-139.  


\bibitem{wang}
Wang SSH, Shultz JR, Burish MJ, Harrison KH, Hof PR, et al (2008) 
Functional trade-offs in white matter axonal scaling. {\it J. Neurosci.} 
{\bf 28}: 4047-4056.

\bibitem{perez}
Perez-Escudero A, de Polavieja GG (2007) Optimally wired subnetwork
determines neuroanatomy of Caenorhabditis elegans. {\it Proc.
Natl. Acad. Sci. USA} {\bf 104}: 17180-17185.

\bibitem{karbowski2001}
Karbowski J (2001) Optimal wiring principle and plateaus
in the degree of separation for cortical neurons. 
{\it Phys. Rev. Lett.} {\bf 86}: 3674-3677. 

\bibitem{stepanyants}
Stepanyants A, Martinez LM, Ferecsko AS, Kisvarday ZF (2009)
The fractions of short- and long-range connections in the visual cortex.
{\it Proc. Natl. Acad. Sci. USA} {\bf 106}: 3555-3560.   


\bibitem{chklovskii2004}
Chklovskii DB, Mel BW, Svoboda K (2004) Cortical rewiring and information
storage. {\it Nature} {\bf 431}: 782-788.

\bibitem{landau}
Landau LD, Lifshitz EM (1960) {\it Mechanics}. Pergamon Press: New York.


\bibitem{mishchenko}
Mishchenko Y, Hu T, Spacek J, Mendenhall J, Harris KM, Chklovskii
DB (2010) Ultrastructural analysis of hippocampal neuropil from
the connectomics perspective. {\it Neuron} {\bf 67}: 1009-1020.   

\bibitem{bar}
Bar TH (1980) The vascular system of the cerebral cortex. 
{\it Adv. Anat. Embryol. Cell Biol.} {\bf 59}: 71-84.    

\bibitem{schmolke}
Schmolke C, Schleicher A (1989) Structural inhomogeneity in the neuropil
of lamina II/III in rabbit visual cortex. {\it Exp. Brain Res.} {\bf 77}:
39-47.   


\bibitem{mainen}
Mainen Z, Sejnowski T (1996) Influence of dendritic structure on firing
patterns in model neocortical neurons. {\it Nature} {\bf 382}: 363-366.

\bibitem{williams}
Williams V, Grossman RG, Edmunds SM (1980) Volume and surface area
estimates of astrocytes in the sensorimotor cortex of the cat.
{\it Neuroscience} {\bf 5}: 1151-1159.  

\bibitem{pawlik}
Pawlik G, Rackl A, Bing RS (1981) Quantitative capillary topography and blood
flow in the cerebral cortex of cat: an in vivo microscopic study. 
{\it Brain Res.} {\bf 208}: 35-58.  

\bibitem{christensen}
Christensen JR, Larsen KB, Lisanby SH, Scalia J, Arango V, et al (2007) 
Neocortical and hippocampal neuron and glial cell numbers in the rhesus monkey. 
{\it Anatomical Record} {\bf 290}: 330-340.  

\bibitem{hao}
Hao J, Papp PR, Leffler AE, Leffler SR, Janssen WG, et al (2006) 
Estrogen alters spine number and morphology in prefrontal
cortex of aged female rhesus monkeys. {\it J. Neurosci.} {\bf 26}: 2571-2578.  

\bibitem{escobar}
Escobar G, Fares T, Stepanyants A (2008) Structural plasticity of circuits
in cortical neuropil. {\it J. Neurosci.} {\bf 28}: 8477-8488.  

\bibitem{medalla}
Medalla M, Barbas H (2009) Synapses with inhibitory neurons differentiate
anterior cingulate from dorsolateral prefrontal pathways associated with
cognitive control. {\it Neuron} {\bf 61}: 609-620.   

\bibitem{dumitriu}
Dumitriu D, Hao J, Hara Y, Kaufmann J, Janssen WG, et al (2010) 
Selective changes in thin spine density and morphology in monkey prefrontal 
cortex correlate with aging-related cognitive impairment.
{\it J. Neurosci.} {\bf 30}: 7507-7515.   

\bibitem{alonso}
Alonso-Nanclares L, Gonzalez-Soriano J, Rodriguez JR, DeFelipe J (2008)
Gender differences in human cortical synaptic density. 
{\it Proc. Natl. Acad. Sci. USA} {\bf 105}: 14615-14619.  

\bibitem{virgintino}
Virgintino D, Monaghan P, Robertson D, Errede M, Bertossi M, et al
(1997) An immunohistochemical and morphometric study on astrocytes
and microvasculature in the human cerebral cortex. {\it Histochemical
Journal} {\bf 29}: 655-660. 

\bibitem{arellano}
Arellano JI, Benavides-Piccione R, Yuste R, DeFelipe J (2007)  
Ultrastructure of dendritic spines: correlation between synaptic and spine
morphologies. {\it Frontiers in Neuroscience} {\bf 1}: 131-143.   

\bibitem{loewenstein}
Loewenstein Y, Kuras A, Rumpel S (2011) Multiplicative dynamics underlie
the emergence of the log-normal distribution of spine sizes in the 
neocortex in vivo. {\it J. Neurosci.} {\bf 31}: 9481-9488.  


\bibitem{song}
Song S, Sj{\"o}str{\"o}m PJ, Reigl M, Nelson SB, Chklovskii DB (2005)
Highly nonrandom features of synaptic connectivity in local cortical
circuits. {\it PLoS Biol.} {\bf 3}: e68.  

\bibitem{peters}
Peters A, Palay SL, Webster DF (1991) {\it The Fine Structure of the Nervous
System}. Oxford: Oxford Univ. Press; 3rd edition.

\bibitem{shepherd}
Shepherd GM, Stepanyants A, Bureau I, Chklovskii D, Svoboda K (2005) 
Geometric and functional organization of cortical circuits.
{\it Nat. Neurosci.} {\bf 8}: 782-790. 

\bibitem{baron}
Baron JC (2001) Perfusion thresholds in human cerebral ischemia: historical
perspective and therapeutic implications. {\it Cerebrovasc. Dis}
{\bf 11} (Suppl. 1): 2-8.   

\bibitem{zhang}
Zhang S, Boyd J, Delaney K, Murphy TH (2005) Rapid reversible changes
in dendritic spine structure in vivo gated by the degree of ischemia.
{\it J. Neurosci.} {\bf 25}: 5333-5338. 

\bibitem{kirov}
Kirov SA, Petrak LJ, Fiala JC, Harris KM (2004) Dendritic spines disappear
with chilling but proliferate excessively upon rewarming of mature
hippocampus. {\it Neuroscience} {\bf 127}: 69-80.   

\bibitem{yasumatsu}
Yasumatsu N, Matsuzaki M, Miyazaki T, Noguchi J, Kasai H (2008)
Principles of long-term dynamics of dendritic spines. {\it J. Neurosci.}
{\bf 28}: 13592-13608. 

\bibitem{tieman}
Tieman SB, Moller S, Tieman DG, White JT (2004) The blood supply of the cat's
visual cortex and its postnatal development. {\it Brain Res.} {\bf 998}: 
100-112.   

\bibitem{oberheim2006}
Oberheim NA, Wang X, Goldman S, Nedergaard M (2006) Astrocytic complexity
distinguishes the human brain. {\it Trends Neurosci.} {\bf 29}: 547-553.  

\bibitem{beardwood}
Beardwood J, Halton JH, Hammersley JM (1959) The shortest path through
many points. {\it Math. Proc. Cambridge} {\bf 55}: 299-327.  

\bibitem{steele}
Steele JM, Shepp LA, Eddy WF (1987) On the number of leaves of an Euclidean
minimal spanning tree. {\it J. Appl. Probab.} {\bf 24}: 809-826.

\bibitem{ogata}
Ogata K, Kosaka T (2002) Structural and quantitative analysis of astrocytes
in the mouse hippocampus. {\it Neuroscience} {\bf 113}: 221-233.  


\bibitem{oberheim2009}
Oberheim NA, Takano T, Han X, He W, Lin JH, et al (2009) 
Uniquely hominid features of adult human astrocytes. 
{\it J. Neurosci.} {\bf 29}: 3276-3287.


\bibitem{boero}
Boero JA, Ascher J, Arregui A, Rovainen C, Woolsey TA (1999) Increased
brain capillaries in chronic hypoxia. {\it J. Appl. Physiol.}
{\bf 86}: 1211-1219.   

\bibitem{meier}
Meier-Ruge W, Hunziker O, Schulz U, Tobler HJ, Schweizer A (1980)
Stereological changes in the capillary network and nerve cells of
the aging human brain. {\it Mechanisms of Ageing and Development}
{\bf 14}: 233-243.  

\bibitem{defelipe}
DeFelipe J, Marco P, Fairen A, Jones EG (1997) Inhibitory synaptogenesis
in mouse somatosensory cortex. {\it Cereb. Cortex} {\bf 7}: 619-634.   

\bibitem{bell}
Bell MA, Ball MJ (1985) Laminar variation in the microvascular architecture
of normal human visual cortex (area 17). {\it Brain Res.} {\bf 335}: 139-143. 

\bibitem{okusky}
O'Kusky J, Collonier M (1982) A laminar analysis of the number of
neurons, glia, and synapses in the visual cortex (area 17) of adult
macaque monkey. {\it J. Comp. Neurol.} {\bf 210}: 278-290.  

\bibitem{huttenlocher}
Huttenlocher PR, Dabholkar AS (1997) Regional differences in synaptogenesis
in human cerebral cortex. {\it J. Comp. Neurol.} {\bf 387}: 167-178.  

\bibitem{erecinska}
Erecinska M, Silver IA (1989) ATP and brain function.
{\it J. Cereb. Blood Flow Metab.} {\bf 9}: 2-19.

\bibitem{press}
Press WH, Teukolsky SA, Vetterling WT, Flannery BP (1992) {\it Numerical
Recipes in Fortran.} Cambridge: Cambridge Univ. Press.

\bibitem{spencer}
Spencer NH (2014) {\it Essentials of Multivariate Data Analysis.} Boca Raton: CRC Press.







\end{thebibliography}
\end{document}